\documentclass[fleqn,usenatbib]{mnras}

\usepackage{newtxtext,newtxmath}

\usepackage{longtable}
\usepackage{geometry}
\usepackage{supertabular}    
\usepackage{threeparttable}
\usepackage{multirow}
\usepackage{xcolor}
\usepackage[utf8]{inputenc}
\DeclareUnicodeCharacter{0301}{\'{e}}
\DeclareUnicodeCharacter{0308}{\'{i}}

\usepackage[T1]{fontenc}
\usepackage{ae,aecompl}
\usepackage{hyperref}

\usepackage{graphicx}	
\usepackage{amsmath}	
\usepackage{amssymb}	
\usepackage{xcolor}

\newcommand{\pbc}{PBC\,J2333.9-2343}

\title[Multiwavelength monitoring of PBC\,J2333.9-2343]{Multiwavelength monitoring of the nucleus in PBC\,J2333.9-2343: the giant radio galaxy with a blazar-like core}

\author[Hern\'andez-Garc\'ia et al.]{
L. Hern\'andez-Garc\'ia$^{1,2}$\thanks{E-mail: lorena.hernandez@uv.cl},
F. Panessa$^{3}$,
G. Bruni$^{3}$,
L. Bassani$^{4}$,
P. Arévalo$^{2,5}$,
\newauthor
V. M. Patiño-Alvarez$^{6,7}$,
A. Tramacere$^{8}$,
P. Lira$^{9,5}$,
P. Sánchez-Sáez$^{10,1}$,
F. E. Bauer$^{11,1,12}$, 
\newauthor
V. Chavushyan$^{6,13}$,
R. Carraro$^{2}$,
F. Förster$^{14,1,15}$,
A. M. Muñoz Arancibia$^{1,15}$,
P. Ubertini$^{3}$
\\
$^{1}$Millennium Institute of Astrophysics (MAS), Nuncio Monseñor Sótero Sanz 100, Providencia, Santiago, Chile \\
$^{2}$ Instituto de F\'isica y Astronom\'ia, Facultad de Ciencias,Universidad de Valpara\'iso, Gran Bretana No. 1111, Playa Ancha, Valpara\'iso, Chile 
		\\
$^{3}$INAF - Istituto di Astrofisica e Planetologia Spaziali, via Fosso del Cavaliere 100, I-00133 Roma, Italy\\
$^{4}$INAF – Osservatorio di Astrofisica e Scienza dello Spazio, via P. Gobetti 101, I-40129 Bologna, Italy \\
$^{5}$Millennium Nucleus on Transversal Research and Technology to Explore Supermassive Black Holes (TITANS) \\
$^{6}$Instituto Nacional de Astrof́ısica, ́Optica y Electŕonica, Luis Enrique Erro 1, Tonantzintla, Puebla 72840, Ḿexico\\
$^{7}$Max-Planck-Institut f̈ur Radioastronomie, Auf dem Hügel 69, D-53121 Bonn, Germany \\
$^{8}$Department of Astronomy, University of Geneva, Ch. d’  ́Écogia 16, Versoix, 1290, Switzerland \\
$^{9}$Departamento de Astronomía, Universidad de Chile, Casilla 36D, Santiago, Chile \\
$^{10}$European Southern Observatory, Karl-Schwarzschild-Str. 2, 85748, Garching,Germany \\
$^{11}$Instituto de Astrof\'isica and Centro de Astroingenier\'ia, Facultad de F\'isica, Pontificia Universidad Cat\'olica de Chile, Casilla 306, Santiago 22, Chile \\
$^{12}$Space Science Institute, 4750 Walnut Street, Suite 205, Boulder, CO 80301, USA \\
$^{13}$Center for Astrophysics | Harvard \& Smithsonian, 60 Garden Street, Cambridge, MA 02138, USA \\
$^{14}$Data and Artificial Intelligence Initiative (IDIA), Faculty of Physical and Mathematical Sciences, University of Chile, Chile \\
$^{15}$Center for Mathematical Modeling (CMM), Universidad de Chile, Beauchef 851, Santiago 8320000, Chile \\
}

\date{Accepted XXX. Received YYY; in original form ZZZ}
 
\pubyear{2023}

\begin{document}
\label{firstpage}
\pagerange{\pageref{firstpage}--\pageref{lastpage}}
\maketitle

\begin{abstract}

 PBC\,J2333.9-2343 is a giant radio galaxy at z = 0.047 with a bright central core associated to a blazar nucleus. If the nuclear blazar jet is a new phase of the jet activity, then the small orientation angle suggest a dramatic change of the jet direction.
We present observations obtained between September 2018 and January 2019 (cadence larger than three days) with Effeslberg, SMARTS-1.3m, ZTF, ATLAS, \emph{Swift}, and Fermi-LAT, and between April-July 2019 (daily cadence) with SMARTS-1.3m and ATLAS. 
Large ($>$2$\times$) flux increases are observed on timescales shorter than a month, which are interpreted as flaring events. 
The cross correlation between the SMARTS-1.3m monitoring in the NIR and optical shows that these data do not show significant time lag within the measured errors.
A comparison of the optical variability properties between non-blazars and blazars AGN shows that \pbc\ has properties more comparable to the latter. The SED of the nucleus shows two peaks, that were fitted with a one zone leptonic model. Our data and modelling shows that the high energy peak is dominated by External Compton from the dusty torus with mild contribution from Inverse Compton from the jet. The derived jet angle of 3 degrees is also typical of a blazar.
Therefore, we confirm the presence of a blazar-like core in the center of this giant radio galaxy, likely a Flat Spectrum Radio Quasar with peculiar properties.
\end{abstract}

\begin{keywords}
galaxies -- individual: PBC\,J2333.9-2343, galaxies -- nuclei,  galaxies -- active
\end{keywords}


\section{Introduction}

Active Galactic Nuclei (AGN) are thought to be powered by supermassive black holes at the center of the galaxies, which are fed by matter falling from the accretion disk \citep{rees1984}. 
Among nearby AGN, only $\sim$10$\%$ show biconical relativistic jets \citep{panessa2016, padovani2017}. Jets are produced by charged
particles accelerated and collimated relativistically in a strong
magnetic field. Jets imprint their mark at radio frequencies due to synchrotron radiation and these AGN are classified as radio galaxies. When one of the jets is oriented close to the line of sight of the observer, its emission is relativistically Doppler boosted, and can dominate over all of the other sources of radiation \citep{blandford1978, giommi2012b}. 
The simplistic Unified Model (UM) of AGN tries to explain the different types of AGN due to orientation effects, with radio galaxies being those where the pair of jets can be observed in the plane of the sky, and blazars represent sources where one of the jets is pointing to the line of sight to the observer \citep{urry1995}.

It is commonly believed that the AGN phenomenon is a phase in the life cycle of a galaxy, where the nuclear activity can be reactivated 10-100 times during its lifetime, with typical timescales for this phases of about 10$^5$-10$^6$ years \citep{hickox2014,schawinski2015,shen2021}. The total growth time of a supermassive black hole has been estimated to be 10$^7$–10$^9$ yr \citep{marconi2004,konar2006}.

One example of the reactivation of AGN activity is from Giant Radio Galaxies \citep[GRGs,][]{ishwara1999,lara2001}, which are characterized by showing linear sizes larger than 0.7 Mpc from radio-lobe to radio-lobe. 
Some GRGs show multiple episodes of nuclear activity that can be observed in the same radio image at different spatial scales. The farthest lobes correspond to the oldest ejection, where the most energetic electrons have already cooled, whereas the structures located closer to the nucleus represent younger and active jets. This is usually interpreted as recurrent nuclear activity, where the old jets are relics of past radio activity \citep[][and references therein]{lara1999, lara2002, saikia2009, gopal2012, bruni2020}. 
A particular case are the X-shaped radio galaxies \citep[XRGs,][]{leahy1984,leahy1992}, which are characterized by exhibiting two misaligned pairs of radio lobes, with the primary lobes created by a pair of powerful radio jets emanating from the central AGN, whereas the origin of the secondary pair is still under debate. Recent works have indeed proposed that not only one scenario can explain their shapes, but that this morphology can be the result of different mechanisms, with one of the options being the reorientation of the jets \citep[see][and references therein]{joshi2019,bhukta2022}.

\pbc\ is a radio galaxy at redshift z = 0.047 with observational characteristics that point to an extreme reinitiated nuclear activity. It was first selected from the INTEGRAL sample because of its different classifications when observed at different wavelengths \citep{basani2016}.
At Mpc scales, the Very Large Array (VLA) radio map shows two jets with a linear size of 1.2 Mpc \citep{basani2016}, whereas at milliarcsecond scale, the Very Long Baseline Array (VLBA) images show an optically thick
compact core with an inverted, self-absorbed spectrum and a steep-spectrum, optically thin jet that extends out asymmetrically to over 60 mas, i.e., $\sim$53 pc in projection.
 Based on the observed radio spectral index, the angle between the jet and the observer must be smaller than 40 degrees \citep{lore2017}. Furthermore, its infrared WISE colors and association with a radio source would classify it as BL Lac \citep{dabrusco2014},
and the modeling of the spectral energy distribution (SED) constrains the angle between the small-scale jet and the observer to less than 6 degrees \citep{lore2017}.
The most plausible explanation for resolving the discrepancy for the different classifications given to this AGN is that, in the past, the nuclear activity was restarted and the jets changed direction so that now a jet is pointing towards us. Indeed, \cite{bruni2020} presented a deep image with the Giant Metrewave Radio Telescope (GMRT) at 150 MHz showing a lack of emission between the lobes and the core region, reinforcing the scenario for which the jet has changed its direction due to a major event, like a galaxy major merger.
 This would imply that for the first time we are observing a transformation from a GRG (with two lobes in the plane of the sky) to a blazar (with a jet pointing towards us), a very exceptional case of jet reorientation. 
Previous observations of PBCJ2333.9-2343 showed variations at all observed wavelengths between radio and X-rays, as well as in the broad optical emission lines, with flux and spectral variations of about 60\%, but the epochs of observation were so sparse ($>$1 year) that the typical timescales could not be estimated \citep{lore2018}. 
In order to constrain the timescales of these variations and to shed light on the nature of this source, we organized a multi-frequency monitoring that was carried out during 2018 and 2019,
where contemporaneous observations were obtained with \emph{Swift} (X-rays and ultraviolet),
SMARTS-1.3m (infrared and optical photometry), and Effelsberg-100m (radio frequency). This study is complemented by data provided through the alert systems Zwicky Transient Facility (ZTF), the Asteroid Terrestrial-impact Last Alert System (ATLAS), plus detections by the Very Long baseline Array (VLBA), the Very Large
Array Sky Survey (VLASS), and the Rapid ASKAP Continuum Survey (RACS), in the radio, and Fermi-LAT at gamma-rays.

The paper is organized as follows. In Section 2 we present the observations and data reduction. In Section 3 the results from the variability analysis, cross correlation and the SED can be found.
A discussion is presented in Section 5. Finally, the main results of this work are summarized in Section 6.

\section{Observations and data reduction}

In this section we present the multiwavelength data that will be used for the analysis. Figures with the light curves for each instrument and Tables containing the data are presented in Appendix A.

\subsection{Radio data}

\subsubsection{Effelsberg-100m monitoring}
We performed a multi-frequency monitoring with the Effelsberg-100m radio telescope, from June to December 2018, with a cadence of almost once per month (project ID 15-18). Secondary focus receivers at four different frequencies were used: 4.8, 8.5, 10.5, and 20.4 GHz (see Table \ref{Effelsberg} for technical details). Observations were performed in cross-scan mode, correcting the pointing on suitable calibrators at a few tens of arcminutes distance from the target. The flux density scale was calibrated on well known sources from \cite{1994A&A...284..331O}. Data reduction was performed with the {\tt{TOOLBOX2}}\footnote{\url{https://gitlab.mpifr-bonn.mpg.de/effelsberg/toolbox2.git}} software, extracting the flux density via single-component Gaussian fit of the cross-scans. The measurement errors were calculated as the sum in quadrature of the cross-scan rms and a 10\% of the total flux density, with the latter being a conservative estimate of the uncertainty on the absolute flux scale calibration.

\subsubsection{Very Long Baseline Array}
Observations with the Very Long Baseline Array (VLBA) were performed on June 28th, 2018, at 15 GHz and 24 GHz (project ID BB390A). The same setup of the previous observations presented in \cite{lore2017} was adopted. The on-source time was 0.5 hours at 15 GHz, while 1.5 hours at 24 GHz. Data were reduced with the astronomical  image  processing  system  ({\tt{AIPS}}\footnote{\url{http://www.aips.nrao.edu}}) following standard procedures. Imaging was performed in {\tt{DIFMAP}}\footnote{\url{https://science.nrao.edu/facilities/vlba/docs/manuals/oss2013a/post-processing-software/difmap}}, via several cycles of phase and phase-amplitude self-calibration. 

\subsubsection{Ancillary data from new radio surveys}
Additionally, we considered data from the Very Large Array Sky Survey at 3 GHz \citep[VLASS,][]{lacy2020} observed on 30 June 2019, and the Rapid ASKAP Continuum Survey at 0.88 GHz (RACS, \citealt{racs}) obtained on 27 March 2020. These data will be used for visualization purposes only (see Section \ref{SEDsect}).

\subsection{Infrared and optical photometry with SMARTS}

\pbc\ was monitored using the dual optical/NIR photometer ANDICAM, that have a pixel scale of 0.37''/px (optical) and 0.27''/px (NIR), mounted on the SMARTS-1.3m telescope, located in Cerro Tololo, Chile. 
We carried out two observing campaigns. The first one used the $V$ and $K$ bands, had a four days average cadence with three exposures per epoch and was carried out between August 2018 and January 2019, to coincide with the targeted campaigns in the other energy bands. The second campaign used the $I$ and $K$ filters and was carried out between April-July 2019 with an average cadence of one day and four exposures per epoch, to refine the measurement of rapid optical and NIR fluctuations. 

The $V$ and $I$ band images were bias-subtracted and corrected by flat-fielded using the observatory standard pipeline. The reduced images were aligned to a reference frame using the {\sc xyxymatch, geomap} and {\sc geotran} {\sc iraf} tasks. The images were degraded to a common seeing of FWHM=7.5 pixels using the task {\sc psfmatch}. We then performed aperture photometry on the target and several other sources in the field using a common coordinate file and fixed small aperture of 8 pixels in diameter, with the task {\sc phot}.

These fluxes were used to build a relative photometry light curve by dividing the flux of \pbc\ by that of the sum of the fluxes of four other sources in the field: 

\begin{equation}
    \frac{ADU_{gal}}{\sum ADU_i}=\frac{f_{gal}}{\sum f_i} \:\Rightarrow \:f_{gal}=\frac{ADU_{gal}}{\sum ADU_i}\cdot \sum f_i
\end{equation}

\noindent where ADU are the counts on the detector and $f_i$ are the calibrated fluxes of the stars obtained from the literature as described at the end of this section.

These comparison sources were selected to be bright but well below the saturation limit and that the pair-wise relative light curves between any pair showed a consistent constant flux ratio. 
Three or four consecutive exposures of 225 s for the $I$ band and 300 s in the $V$ band were taken each observing night. We averaged the relative fluxes within a night and considered the root-mean-squared (RMS) scatter of the fluxes as an estimate of the uncertainty in the flux measurement.  

In the case of the NIR detector, two co-added exposures of 90 s each were taken at three different pointings with a small offset respect to the field size each observing night. The three pointings ensure that the target and two nearby comparison objects appear in each exposure at a different location in the detector, so that consecutive images can be used to subtract the bright sky emission. Naming the exposures as A, B and C in chronological order, we computed the difference images A-B, B-A and C-B using the task {\sc imarith} in {\sc iraf} to produce three sky-subtracted images. We further degraded these to a common FWHM of 8 pixels with the task {\sc gauss} in {\sc iraf} and extracted aperture photometry for the targets and two comparison objects with an aperture diameter of 8 px. As the field of view of the NIR detector is smaller than that of the optical detector, we could only use the nearest two comparison objects of the four used in the optical case. One of them, however, was too dim to produce useful photometric data so the final relative photometry was done with respect to the brightest of the two comparison objects, which is 20 times brighter than our target. The three relative photometry points per observing night were averaged together and the RMS scatter between the three points is used as an estimate of the uncertainty.

We then calibrated the relative flux by multiplying it by the sum of the fluxes of the comparison objects, where the reference fluxes in the V and K bands were obtained from the "TESS input catalog v8.0" \citep{stassun2019}, and for the I band it was obtained from the "The USNO-B1.0" \citep{monet2003}.

\subsection{Optical photometry}

\subsubsection{Zwicky Transient Facility (ZTF) \label{ztf}}

The Zwicky Transient Facility \citep[ZTF,][]{graham2019, bellm2019, masci2019} surveys the extragalactic Northern Sky every three days in the g, r and i optical filters since 2018. 
It uses the Palomar Observatory’s Samuel Oschin 48" Schmidt telescope, that has a 47 deg$^2$ field of view (FoV) composed by 16 CCDs and reaches 20.5 r-band mag in 30 seconds exposure time. The images are processed by the Infrared Processing and Analysis Center (IPAC) pipeline. 

ZTF offers different services, including 1) a public alert system, for real-time, time-domain science,
2) Data Releases (DRs) every two months, including photometry measurements on the science images 
and 
3) the Forced Photometry Service on-demand and per source, including photometry measurements on the reference-subtracted science images.                                                                        

For an alert to be generated, a source has to show a variation above 5$\sigma$ of confidence level with respect to a reference image\footnote{A reference image is generated by ZTF from the stacking of at least 10 images of the source.}. 
We cross-matched the coordinates of \pbc\ with the ZTF alert stream using the Web Interface\footnote{\url{https://alerce.online/}} provided by the Automatic Learning for the Rapid Classification of Events \citep[ALeRCE,][]{forster2021} broker, and found that this source (ZTF name is \href{https://alerce.online/object/ZTF18abwpdny}{ZTF18abwpdny}) had alerts during the monitoring that we performed \citep[AT 2018igu,][]{alert}.

For this work we retrieved data from the ZTF Forced Photometry Service. The measurements obtained from this service are less affected by extranuclear emission than in DRs because it is obtained from the reference-subtracted images, therefore isolating the variable nuclear component.
The light curves were cleaned with the following criteria: the processing summary/QA bits for science image \textsc{infobits}=0, we used data only from the CCD with the largest number of data points per filter, as well as rejecting bad-data quality flags that are explained in the ZTF Public DRs\footnote{\url{http://web.ipac.caltech.edu/staff/fmasci/ztf/extended\_cautionary\_notes.pdf}}. The following are filter dependent and are related to the photometric zero point (ZP) for science image (\textsc{zpmaginpsci}, in [mag]), and its deviation from the average (rms) difference between instrumental magnitudes and PanSTARRS1 calibrators (\textsc{zpmaginpscirms}, in [mag]), so we rejected data points that fulfill the following criteria:

{\begin{itemize}
    \item Filter g:
    
        \textsc{zpmaginpsci} > 26.7 - 0.2secz OR
    
         \textsc{zpmaginpsci} < ZP$_{thres}$[rcid] - 0.2secz OR
         
         \textsc{nps1matches} < 80 OR
         
         \textsc{zpmaginpscirms} > 0.06
         
    \item Filter r:
    
        \textsc{zpmaginpsci} > 26.65 - 0.15secz OR
    
         \textsc{zpmaginpsci} < ZP$_{thres}$[rcid] - 0.15secz OR
         
         \textsc{nps1matches} < 120 OR
         
         \textsc{zpmaginpscirms} > 0.05  
    
    \item Filter i:
    
         \textsc{zpmaginpsci} > 26.0 - 0.07secz OR
     
         \textsc{zpmaginpsci} < ZP$_{thres}$[rcid] - 0.07secz OR
         
         \textsc{nps1matches} < 100 OR
         
         \textsc{zpmaginpscirms} > 0.06     
    
   \end{itemize}} 
    
\noindent where secz is the airmass, that was computed by using the EarthLocation function within astropy \citep{astropy:2013, astropy:2018}; ZP$_{thres}$[rcid] is the CCD-quadrant-based ZP thresholds to identify bad quality images\footnote{\url{http://web.ipac.caltech.edu/staff/fmasci/ztf/zp\_thresholds\_quadID.txt}}; and \textsc{nps1matches} is the number of PS1 calibrators. 

The difference PSF magnitudes were converted into apparent magnitude following \cite{forster2021}. The resulting magnitudes were color corrected using the linear color coefficient from ZTF calibration and the colors in Pan-STARRS1 (clrcoeff*(g-r) for g and r, and clrcoeff*(r-i) for the i-filter.
We used data between July-December 2018. 
We note that data during the second monitoring in 2019 is not included because even if there was coverage later on with ZTF, there are only a few data points and do not provide additional information.

\subsubsection{Asteroid Terrestrial-impact Last Alert System (ATLAS)}

The Asteroid Terrestrial-impact Last Alert System \citep[ATLAS,][]{tonry2018} surveys the entire north sky\footnote{ATLAS started monitoring the southern sky in 2022       .} (dec $>$ -50 degrees) every two nights in two optical filters, c (4157.33-6556.44 \AA) and o (5582.07-8249.18 \AA). 
ATLAS began observations in 2015 with one telescope and its two-telescopes version has been operational since 2017. Each of these are 50-centimeter diameter f/2 Wright-Schmidt telescopes, with a 7.4 deg FOV. At 30 seconds per exposure, they reach a magnitude limit of 19.

We obtained the data from the ATLAS Forced Photometry Service\footnote{\url{https://fallingstar-data.com/forcedphot/}} 
on the science images. In order to remove bad data points, we applied filters to the errors and the limiting magnitude as follows. We obtained the cumulative distributions of the flux error and limiting magnitude and determined the percentiles at which the slope of the distributions changed. This resulted in a 90\% percentile for the flux error (\textsc{duJy} $<$ 62) and 5\% percentile for the limiting magnitude (\textsc{mag5sig} $<$ 17.2). We also filtered by date to include only data during our monitoring campaign (MJD between 58320 and 58700).


\subsection{UV data}

The Ultraviolet and Optical Telescope \citep[UVOT,][]{2005SSRv..120...95R} onboard the Neil Gehrels \emph{Swift} Observatory has six primary photometric filters.
During the monitoring only the UVM2 (centered at 2246 \AA) filter was used, in order to maximize the cadence and S/N of the resulting light curve. These observations were performed simultaneously to the X-ray observations weekly between August-November 2018.

The {\sc uvotsource} task within the software HEASoft version 6.26 was used to perform aperture photometry using a circular aperture radius of 5 arcsec centred on the coordinates of \pbc . The background region was selected free of sources adopting a circular region of 20 arcsec close to the nucleus.

\subsection{X-ray data}

We performed weekly monitoring with the \emph{Swift} X-ray Telescope \citep[XRT,][]{2005SSRv..120..165B} onboard the Neil Gehrels \emph{Swift} Observatory in the Photon Counting mode between August-November 2018, with a total of 17 observations taken simultaneously to the UV observations.

The data reduction  was performed following standard routines described by the UK Swift Science Data Centre (UKSSDC) using the software in HEASoft version 6.26. Calibrated event files were produced using the routine {\sc xrtpipeline}, accounting for bad pixels and effects of vignetting, and exposure maps were also created. Source and background spectra were extracted from circular regions with 30 arcsec and 80 arcsec radius, respectively. The {\sc xrtmkarftask} was used to create the corresponding ancillary response files. The response matrix files were obtained from the HEASARC CALibration DataBase. The spectra were grouped to have a minimum of 20 counts per bin using the {\sc grppha} task.

The X-ray data analysis was performed using XSPEC v. 12.10.1. We assumed a Galactic absorption of $N_{Gal} = 1.63 \times 10^{20} cm^{-2}$ \citep{1990ARAA..28..215D}. Following the results obtained in \cite{lore2017} and \cite{lore2018}, where \emph{Swift} data between 2010 and 2017 were analyzed, we fitted the {\it Swift}/XRT spectra with a single power-law model. All the spectra were fitted simultaneously to search for X-ray spectral and flux variability \citep{lore2013}. 
Modeling all epochs with linked photon index, $\Gamma$, and normalizations of the power law (i.e. assuming the non-variable case) results in a  $\chi^2$ of 2481.7 for 1242 degrees of freedom (dof), i.e. a reduced $\chi^2$ of 2).
Allowing the normalizations to vary independently results in a satisfactory fit with $\chi^2$/dof = 1198.08/1121 = 0.99, whereas allowing $\Gamma$ to vary results in a reduced $\chi^2$ of 1.74. Varying together the normalizations and $\Gamma$ did not improve the fit (F-test=0.02).
 From this spectral fit we obtained the intrinsic X-ray luminosity of the source for each epoch.
 
We used \emph{XMM-Newton} data (from pn and OM) analyzed in \cite{lore2017} to compare multiwavelength data of the source in different epochs through its spectral energy distribution (SED, see Section \ref{SEDsect}). We refer the reader to this paper for details on the data reduction and analysis. The only difference between the data presented here, and the one published before, is that there was an error in the unit conversion of a factor two in the X-ray data that is now corrected for a proper comparison.

\subsection{Gamma-Ray Data}
\label{gammarays}

We used data from the Fermi Large Area Telescope \citep[LAT,][]{abdo2009} database. First, we checked that the 95\% error ellipse of this source includes only the nuclear region, excluding the lobes as gamma-ray emitters.
We calculated the gamma-ray flux integrating data between May 1st 2018 at 00:00 (UT) and September 30th 2019 at 24:00 (UT), in order to match the observation epochs of the data in the other wavebands. We considered data in the energy range 0.1–300 GeV, and analyzed it with the Fermitools version 1.0.2. From the 4FGL catalog \citep{abdollahi2020}, all sources within 15° of the location of PBC J2333.9-2343 were included in the model. The spectral parameters for the sources within 5° were left free, while for the remaining sources only the normalization was left free. We note that we were not able to obtain a light curve because the smaller time bins tested did not have enough signal to noise ratio to consider them statistically significant detections (test statistic, TS $\geq$ 25).

For the aforementioned epochs, we obtained an integrated flux of (1140.8$\,\pm\,$6.2)$\,\times\,$10$^{-11}$ photons s$^{-1}$cm$^{-2}$ $\equiv$ (599.4$\,\pm\,$6.5)$\,\times\,$10$^{-14}$ erg s$^{-1}$cm$^{-2}$ at a 6.2$\sigma$ of significance level. The data were fitted with a power law model with spectral index $\Gamma$ = 2.417$\pm$0.002. 

In order to compare with the results in \cite{lore2017}, we performed the same analysis in Fermi-LAT data between January 1st 2015 at 00:00 (UT) and December 31 2015 at 24:00 (UT). We obtained an integrated flux of (603.8$\,\pm\,$3.4)$\,\times\,$10$^{-11}$ photons s$^{-1}$cm$^{-2}$ $\equiv$ (520.0$\,\pm\,$5.9)$\,\times\,$10$^{-14}$ erg s$^{-1}$cm$^{-2}$ at a 5.1$\sigma$ of significance level, and $\Gamma$ = 2.149$\pm$0.002.  

Regarding the use of gamma-ray data to constrain the SED model, we prefer to not use the fluxes obtained above, because they span over three orders of magnitude in energy. Therefore, we made three equal logarithmic energy bins for the same epochs detailed above, to try and build a low-resolution gamma-ray spectra. The three energy bins are between the energies 100 MeV - 1.44 GeV - 20.8 GeV - 300 GeV.

For the 2018-2019 data, the first energy bin resulted in a TS of 19.2, the second energy bin yielded a TS of 14.9, and the third energy bin resulted in a TS lower than 0, which means that the isotropic and diffuse background are brighter than the source at those energies. Since the TS for any of the three bins is not enough to be considered a detection (i.e., TS $\geq$ 25), we decided to compute upper limits (with Bayesian analysis, using FermiPy) for the first and second energy bins -- it was impossible for the third one. For the first energy bin, we obtained an upper limit of 7.8 $\times$ 10$^{-9}$ ph s$^{-1}$ cm$^{-2}$ $\equiv$ 3.2 $\times$ 10$^{-12}$ erg s$^{-1}$ cm$^{-2}$. For the second energy bin, we obtained an upper limit of 2.1 $\times$ 10$^{-10}$ ph s$^{-1}$ cm$^{-2}$ $\equiv$ 1.3 $\times$ 10$^{-12}$ erg s$^{-1}$ cm$^{-2}$.

For the 2015 data, we did not get a significant detection, the TS we obtained for the first, second and third energy bins are 7.5, 20.4 and lower than zero, respectively. This time, we were only able to get an upper limit for the second energy bin, for the third energy bin it was impossible, and for the first one, the analysis did not converged satisfactorily. For the second energy bin, we obtained an upper limit of 3.7 $\times$ 10$^{-10}$ ph s$^{-1}$ cm$^{-2}$ $\equiv$ 1.6 $\times$ 10$^{-12}$ erg s$^{-1}$ cm$^{-2}$. These upper limits will be used in Section \ref{SEDsect}.

In order to visually compare our upper limits, we included the results reported in \cite{abdollahi2020}, who presented the Data Release 3 (4FGL-DR3) for Fermi-LAT including 12 years of survey data. PBC\,J2333.9-2343 shows detections in three out of the eight energy bands: 0.3-1, 1-3, and 3-10 MeV (see Section \ref{SEDsect}).


\begin{figure*}
	\includegraphics[width=17cm]{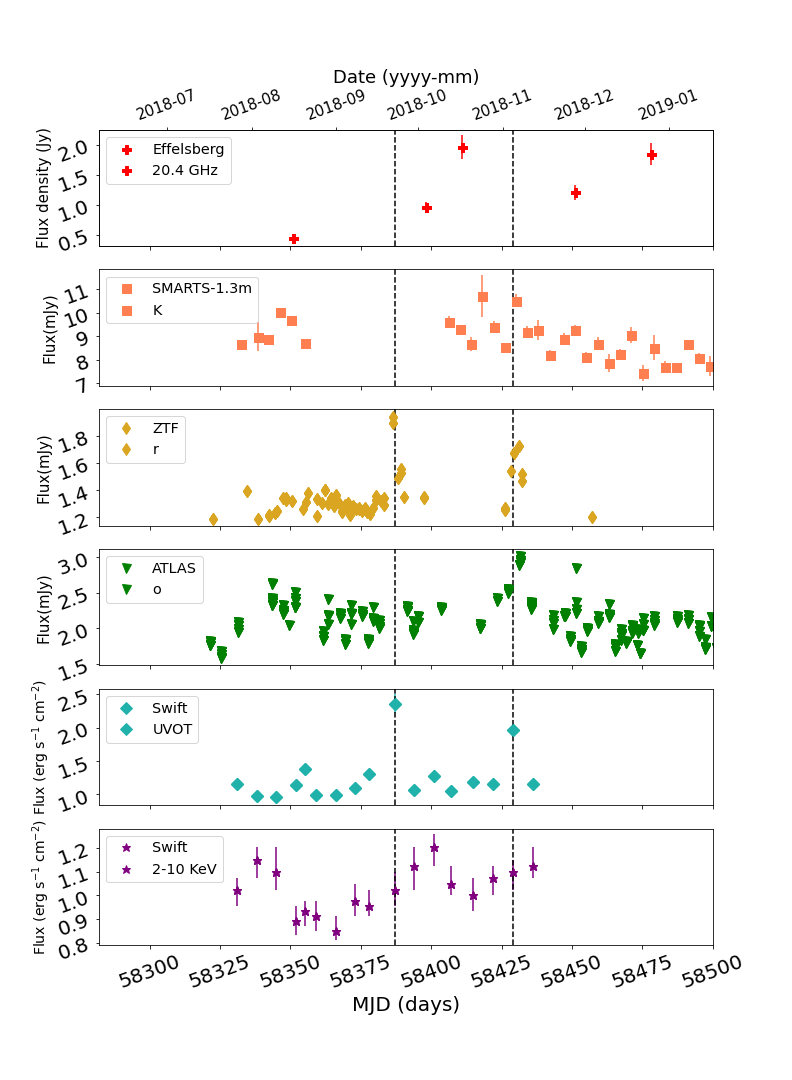}
\caption{Multiwavelength light curve of the monitoring of \pbc\ during 2018. Note that for more clarity, only one light curve per instrument is plotted here. The grey dashed lines represent the peak of the first and second flares as observed by \emph{Swift}/UVOT/UVM2 that occurred at MJD=58387 and 58429.}
    \label{fig:multilambda}
\end{figure*}


\section{Results}

\subsection{Variability analysis}
\label{variability}

\begin{figure*}
	\includegraphics[width=11cm]{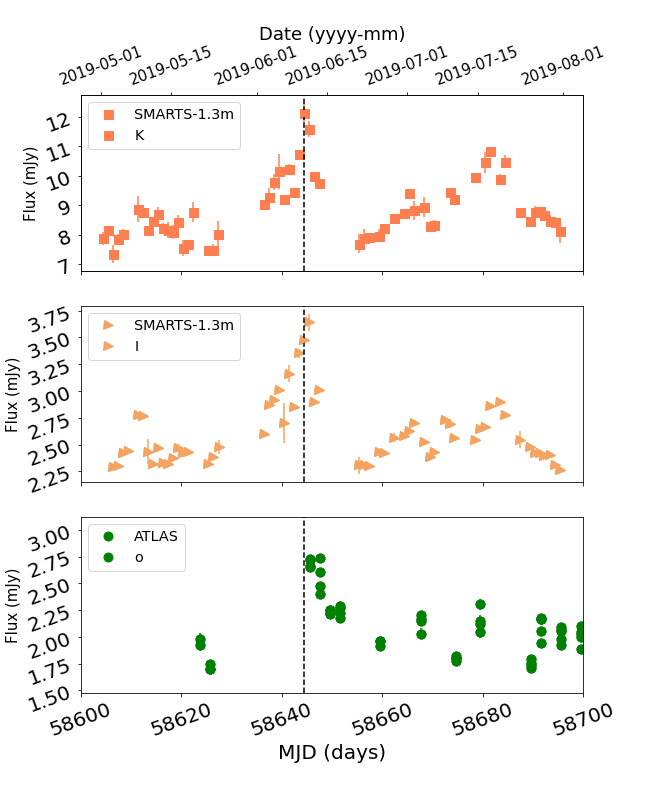}
\caption{Multiwavelength light curve of the monitoring of \pbc\ during 2019. Observations are only available with SMARTS-1.3m and ATLAS. The grey dashed lines represent the peak of the third flare at MJD=58644.5.}
    \label{fig:tercerpeak}
\end{figure*}

In Fig. \ref{fig:multilambda} we plot the multiwavelength light curves for the monitoring in 2018, spanning between MJD=58320 (July 21st, 2018) and 58500 (January 17th, 2019). For simplicity and clarity, we plot only the most representative light curve per instrument (see Appendix A for light curves in all bands). The selection of the light curves obeys to the following criteria: for Effelsberg we selected the 20.4 GHz because is the only one showing variations in the radio, from SMARTS-1.3 the K band  shows NIR variations, from ZTF and ATLAS the r-band and o-band because these have better cadence, from \emph{Swift}/UVOT there is only one band, and from \emph{Swift}/XRT the 2-10 keV band because it represents the nuclear source.
In Fig. \ref{fig:tercerpeak} we plot the monitoring performed between MJD=58600-58700 (April 27th-August 5th, 2019) with SMARTS-1.3m and also observed by ATLAS. These observations were taken with a cadence of 1-3 days, allowing a more detailed analysis of the observed variations.


We show the results of the variability analysis of these light curves in Table \ref{tab:variability}. For each of the observed bands, we list the mean flux and its standard deviation. The $\chi^2$ of the light curve with respect to a constant flux is also presented with the degrees of freedom (d.o.f). 
In addition, we have estimated the normalized excess variance, $\sigma^{2}_{NXS}$ and its error $\Delta \sigma^{2}_{NXS}$, which represents the variability amplitude of the light curves, following the prescriptions in \cite{vaughan2003}. A source is considered to be variable when $\sigma^{2}_{NXS}$ $>0$ within the errors, i.e. when the intrinsic amplitude of the variability is greater than zero. 
We also report F$_{var}$, i.e., the intrinsic variability amplitude as in  \cite{vaughan2003}.
Finally we include the percentage of the variations, estimated as the change between the minimum and maximum flux in the light curves.

The results obtained from this analysis are the following:

\begin{itemize}
    \item Flux variations are found in almost all the observed frequencies:
    \begin{itemize}
         \item In the radio, flux densities at different epochs with Effelsberg at 4.8, 8.5, and 10.5 GHz are consistent within errors, thus showing no variability. At 20 GHz, a significant difference between epochs is found, with values varying by a factor of four - between $\sim$0.5 Jy and $\sim$2 Jy - in only two months, at a 13$\sigma$ of confidence level. The fact that variability is only evident at frequencies $>$10 GHz is most probably due to the different radio-emitting regions probed at high frequencies, the latter being linked to the inner part of the jet (closer to the core) and thus more subject to Doppler boosting. 
         \item Variations in the K band flux observed by SMARTS-1.3m between 7.3 and 12.13 mJy are detected, at a 6$\sigma$ of confidence level in 2018, and 20$\sigma$ during the 2019 monitoring. 
         \item Optical variations in the g,r,i (ZTF), o,c (ATLAS), and $V$,$I$ (SMARTS-1.3m) by a factor of about two are observed in all bands. In all cases these variations are detected at confidence levels larger than 14$\sigma$.
         \item The UV \emph{Swift} observations show changes between 0.96 and 2.36 $\times$ 10$^{-15}$ erg s$^{-1}$cm$^{-2}$, i.e., a flux increase by a factor of 2.5. The variations are detected at 16 $\sigma$ of confidence level. 
         \item X-ray variations are detected only at a 2$\sigma$ of confidence level, with flux variations by a factor of 1.4. 
    \end{itemize}
    \item The observed flux variations show a flaring behaviour. Three events occurred during this monitoring and were detected at different frequencies:
    \begin{itemize}
        \item On September 25th, 2018 the ZTF reported a first alert coming from the coordinates of PBC\,J2333.9-2343. These alerts are produced only for changes above 5$\sigma$ when compared to a reference image. 
        This first flare had its peak around September 26th, 2018 (MJD=58387) and was detected by \emph{Swift}/UVOT, and ZTF (see Fig. \ref{fig:multilambda}).
        The data points before and after this date observed by \emph{Swift}/UVOT, when the higher amplitude of the variation is detected, were September 19th and October 3rd, 2018, i.e., 16 days, so this represents the maximum duration for this first flare. The left black dashed line of Fig. \ref{fig:multilambda} represents the highest point in \emph{Swift}/UVOT.
        \item The second flare peaked around November 11th, 2018 (MJD=58429) and was detected by \emph{Swift}/UVOT, ZTF, ATLAS, and SMARTS-1.3m. The data points before and after this date observed by \emph{Swift}/UVOT were October 31st and November 14th, 2018, i.e., 15 days. The right black dashed line of Fig. \ref{fig:multilambda} represents the highest point in \emph{Swift}/UVOT during this flare.
        \item The third flare peaked on June 10th, 2019 (MJD=58644.5). This was detected by SMARTS-1.3m and ATLAS, the only telescopes that were monitoring during that period. The flare occurred between May 25th and June 22nd, 2018 as observed by SMARTS-1.3m, i.e., 28 days (see Fig. \ref{fig:tercerpeak}).      \item Effelsberg also shows variations at 20.4 GHz, but the sampling of the light curve does not allow a detailed study at this frequency. The maximum flux observed at 20.4 GHz was on November 20th, 2018, i.e., in the middle of the first and second flare, so we cannot assume that it is related to any of the flares.
    \end{itemize}
    \item In order to quantify the variability timescales of the source, we calculated the doubling/halving times, $t_d$ \citep[see e.g.][]{Brown2013, Saito2013, Kapanadze2018, Abhir2021}, using the SMARTS-1.3m light curves from 2019, which have the highest cadence among our data sets. The doubling/halving times represent how long it will take for a variable time series, to double (if it increases) or halve (if it decreases) its flux, assuming the variability pattern (in this case, a power-law function) does not change during the time period of interest (see Appendix~\ref{appendixb} for details). The shortest $t_d$ in the I band is 7.7 days, and in the K band is 6.7 days. Using these timescales we computed the size of the emission region following the relation $R \leq ct_{var} \delta_D/(1+z)$ \citep{abdo2011}. We estimated the Doppler factor $\delta_D = [\Gamma(1- \beta cos \theta)]^{-1}$=2.7 using $\theta$=3 degrees and $\Gamma$ and $\beta$ from \cite{lore2017}. Assuming $t_{var}$=$t_d$, we obtain a region of 5.1$\times$10$^{16}$ cm in the I band, and of 4.4$\times$10$^{16}$ cm in the K band.  
\end{itemize}

\begin{table*}
\centering
\caption{Results of the variability analysis. For the 2018 (Fig. \ref{fig:multilambda}) and 2019 (Fig. \ref{fig:tercerpeak}) monitorings, and for each band, it lists the mean flux and its standard deviation (in mJy), the value of $\chi^2$ and the degrees of freedom, the normalized excess variance, $\sigma^2_{NXS}$, and its error, the intrinsic variability amplitude, $F_{var}$ and its error, and the percentage of variation.  \label{tab:variability}}
\begin{tabular}{lcccccc}
\hline
Band & Mean (mJy)	&	Stddev (mJy)	&	$\chi^2$	/	d.o.f	&	$\sigma^2_{NXS} \pm \Delta \sigma^2_{NXS}$	&	$F_{var} \pm \Delta F_{var}$	&	Change ($\%$)	\\ \hline
\multicolumn{7}{c}{Monitoring 2018} \\
\hline
Effelsberg-4.8 & 1156.0	&	43.3	&	0.8	/	6	&	<0	&		-	&	10.4	\\ 
Effelsberg-8.5 & 1331.3	&	65.7	&	1.7	/	6	& <0	&		-	&	13.0	\\ 
Effelsberg-10.5 & 1456.3	&	98.0	&	3.0	/	6	&	<0	&		-	&	16.8	\\ 
Effelsberg-20.4 & 1122.0	&	774.1	&	247.7/2 		&	0.47 	$\pm 	$0.08	&	0.68 	$\pm 	$0.04	&	77.4	\\ 
SMARTS-V & 1.2 	& 	0.1 		&	7189.3	/	24	&	0.0079	$\pm	$0.0004	&	0.0886	$\pm	$0.0002	&	32.5	\\ 
SMARTS-K & 9.0 	& 	0.8	&	4375.2	/	24	&	0.006	$\pm	$0.001	&	0.0793	$\pm	$0.0005	&	30.5	\\ 
ZTF-g & 0.70	&	0.08	&	12045.7	/	45	&	0.0145	$\pm	$0.0003	&	0.1204	$\pm	$0.0002	&	45.9	\\ 
ZTF-r & 1.3	&	0.2	&	23291.6	/	62	&	0.0135	$\pm	$0.0002	&	0.11615	$\pm	$0.00008	&	38.9	\\ 
ZTF-i & 1.8	&	0.1	&	1320.2	/	10	&	0.0067	$\pm	$0.0003	&	0.0819	$\pm	$0.0001	&	24.3	\\ 
ATLAS-o & 2.1	&	0.3	&	20554.3	/	165	&	0.0167	$\pm	$0.0003	&	0.1293	$\pm	$0.0001	&	47.6	\\ 
ATLAS-c &1.2	&	0.1	&	1050.7	/	33	&	0.0069	$\pm	$0.0005	&	0.0828	$\pm	$0.0003	&	27.8	\\ 
\emph{Swift}-UVM2 & 0.00009 	& 	0.00003 	&	514.9 /	16	&	 0.086	$\pm	$0.006  &	 0.296$\pm	$0.003  &	59.2	\\ 
emph{Swift}-(0.5-2  keV) & 0.0018	&	0.0002	&	49.4	/	16	&	0.005	$\pm	$0.002	&	0.073	$\pm	$0.001	&	27.6	\\ 
\emph{Swift}-(2-10  keV) & 0.00071 	& 	0.00007 	& 	42.8 	/ 	16 	& 	0.005 	$\pm 	$0.002 	& 	0.071 	$\pm 	$0.001 	& 	29.2	\\
\hline
\multicolumn{7}{c}{Monitoring 2019} \\
\hline
SMARTS-I & 2.6	&	0.3	&	20064.4	/	58	&	0.0132	$\pm	$0.0003	&	0.1150	$\pm$ 0.0002	&	37.8	\\ 
SMARTS-K & 8.9	&	1.0	&	3648.7	/	58	&	0.0132	$\pm	$0.0007	&	0.1150	$\pm$ 0.0003	&	39.5	\\ 
ATLAS-o & 2.1	&	0.3	&	6927.1	/	54	&	0.0169	$\pm	$0.0005	&	0.1299	$\pm$ 0.0003	&	37.9 	\\ 
\hline
\end{tabular}
\end{table*}

\subsection{Cross-Correlation analysis}

The monitoring in 2018 had roughly weekly cadence, thus it is not possible to evaluate the similarity between the light curves. For this reason, we did not perform the cross-correlation analysis on these data. We will use the 2019 SMARTS-1.3m monitoring, which has cadence of 1-3 days and simultaneous observations in both bands, to cross-correlate the light curves in the K and I bands.

\begin{figure}
	\includegraphics[width=0.5\textwidth]{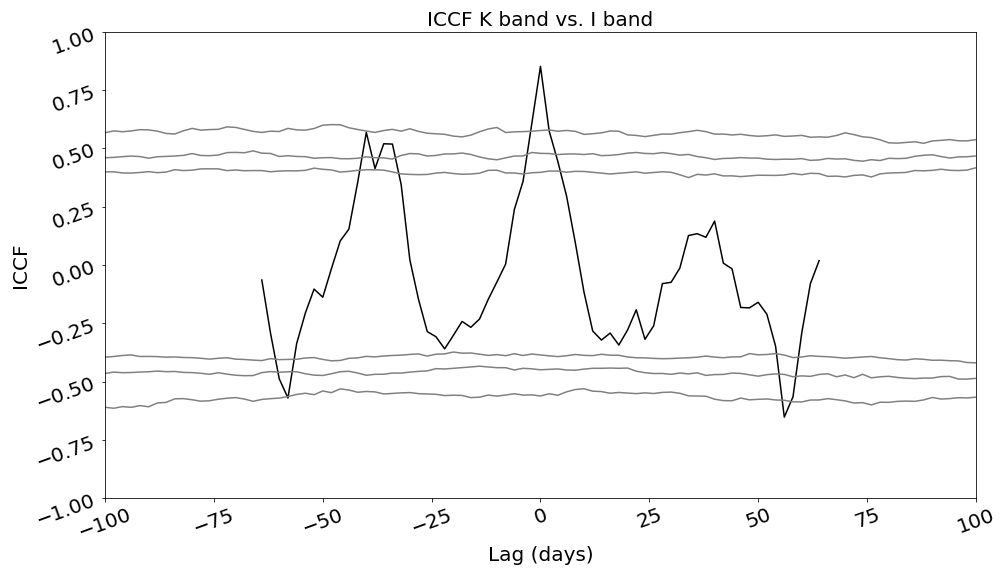}
\caption{The cross correlation function obtained by the interpolation method using SMARTS-1.3m data. The noisy horizontal lines represent the 90, 95 and 99\% significance, both at correlation (above correlation coefficient above 0) and anti-correlation (below 0).}
    \label{ccfigure}
\end{figure}

The cross-correlation function (CCF) was estimated using three methods: the Interpolated Cross Correlation Function \citep[ICCF,][]{iccf1986}, Discrete Cross-Correlation Function \citep[DCF,][]{dcf1988}, and the Z-Transformed Discrete Cross-Correlation Function \citep[ZDCF][]{alexander1997}. We applied these methods with the modifications to be used for non-stationary time series following \cite{patino2018} and \cite{almazan2022}. We also calculated significance levels, following \cite{emmanoulopoulos2013}. The cross-correlation function between the K and I bands is shown in Figure~\ref{ccfigure}. We consider a significant delay only when it is above 99\% significance and obtained in at least two of the methods. We obtained a delay by averaging the lag obtained with the three methods, of 1.02$\pm$1.45 days, i.e., the cross-correlation analysis did not find either delay or advance within the measurement errors, and is compatible with these variations occurring quasi-simultaneously.

\subsection{\label{SEDsect}Spectral energy distribution (SED)}

\begin{figure*}
	\includegraphics[width=7.84cm]{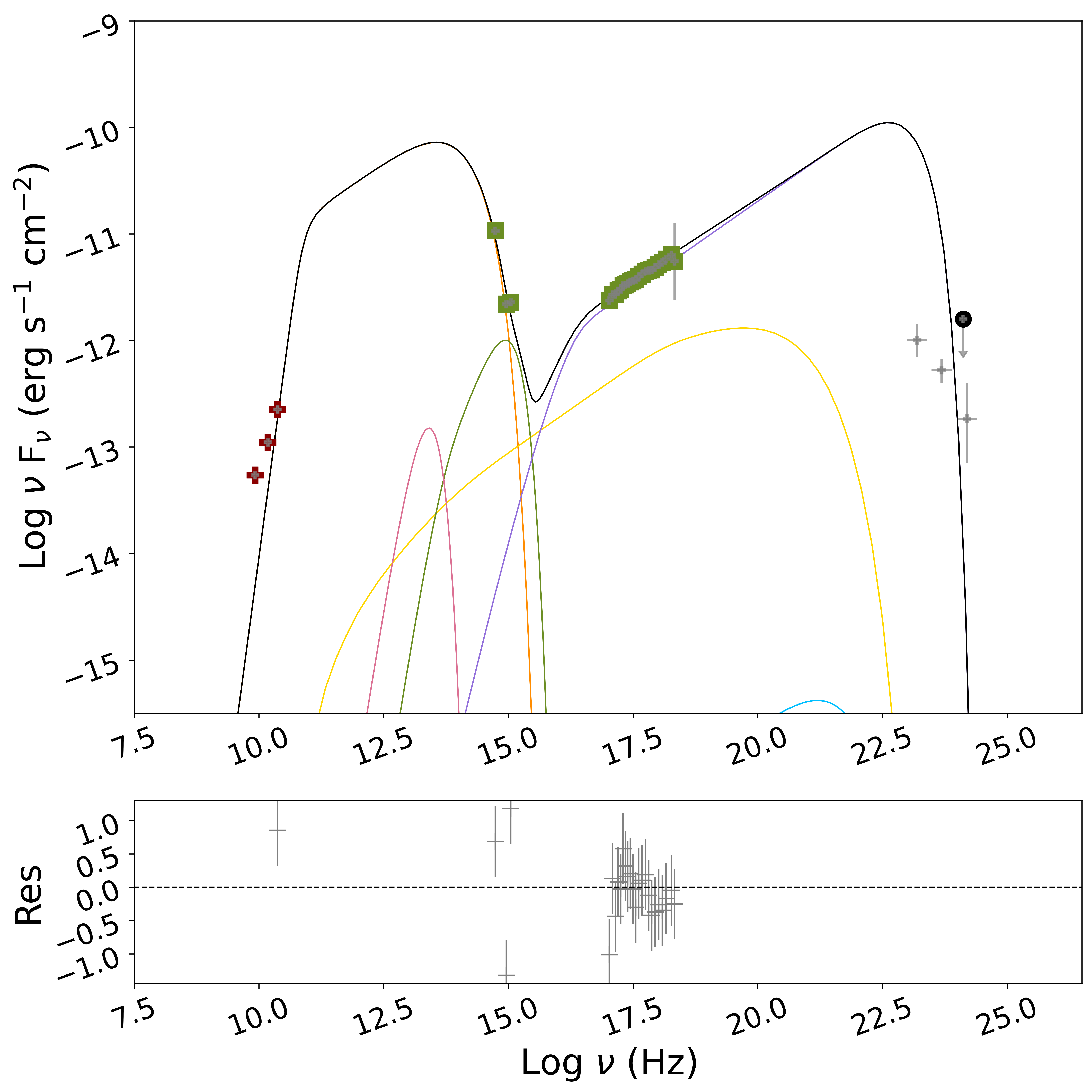}
 \includegraphics[width=9.76cm]{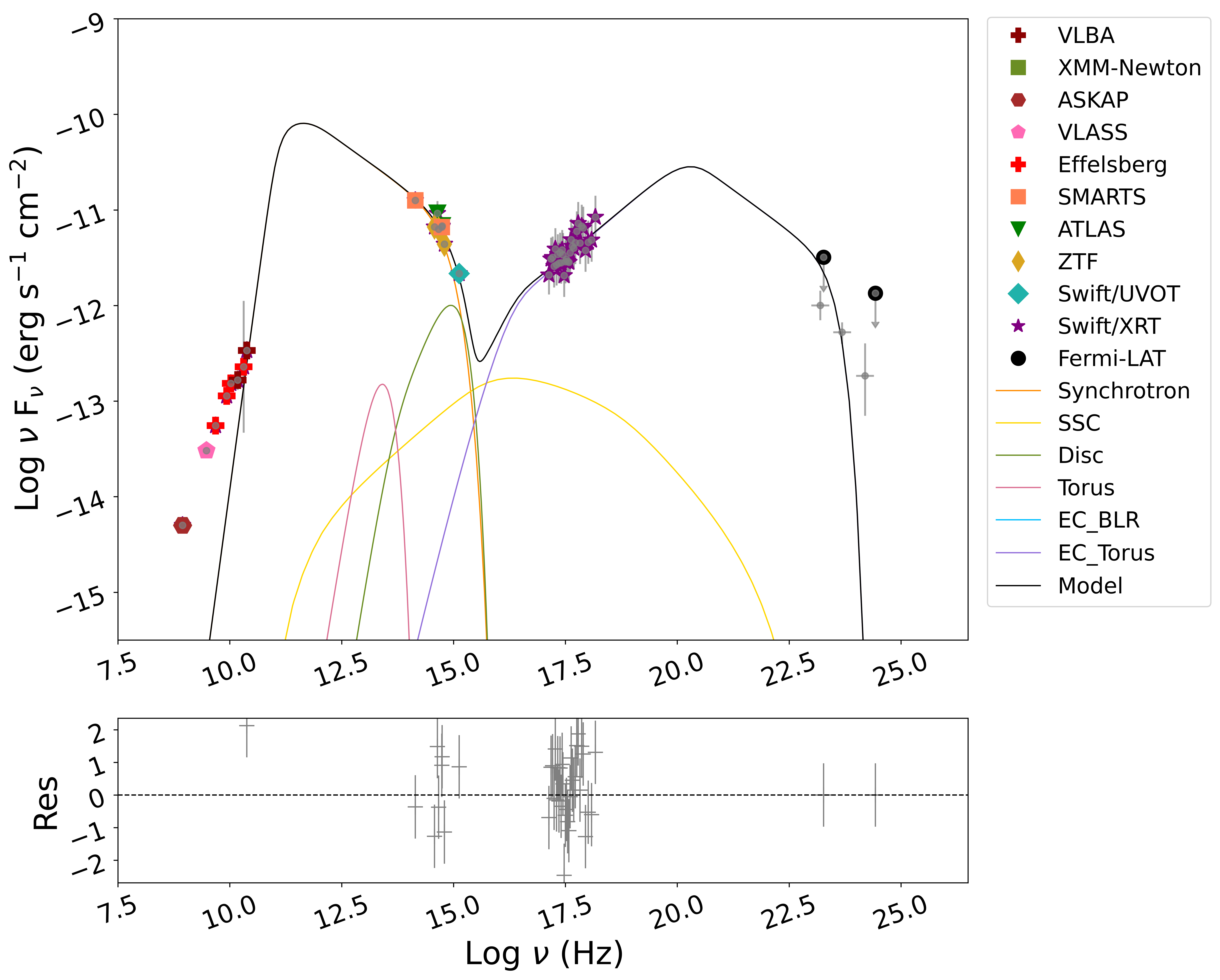}
\caption{Left: The observed frame SED of PBC\,J2333.9–02343 built with data from 2015 (as in \citealt{lore2017}), including data from VLBA, \emph{XMM-Newton} and Fermi-LAT (not included in previous works).
Right: The observed frame SED using contemporaneous data from 2018 from
the Effelsberg, SMARTS-1.3m, ZTF, ATLAS, and \emph{Swift} observatories. Additionally we included data from VLASS, RACS (only for visualization), VLBA and Fermi-LAT. The fit was done using the upper limits from Fermi-LAT, but for visualization we also include the 4FGL-DR3 detections.}. The solid lines represent the different components of the fitted model: the orange line represents the synchrotron emission, the yellow line represents the synchrotron self-Compton (SSC), the pink line represents the dusty torus, the purple line represents the inverse Compton of the dusty torus, the green line represents the accretion disk, the blue line represents the inverse Compton from the broad line region and the black line is the sum of all the aforementioned components. The bottom panels include the residuals, which correspond to data with systematics. 
    \label{fig:sed}
\end{figure*}

In order to construct the SED, we used data obtained when there was no flaring activity in the different bands, i.e., before MJD = 58370. When more than one data point was available, we computed the mean value. The SED is presented in the right panel of Fig. \ref{fig:sed}. We included RACS and VLASS data for visualization purposes, but data below 20 GHz were not considered because 
compact regions produce a synchrotron spectrum that is self–absorbed, and therefore cannot account for the radio flux at smaller frequencies. This has to be
produced by other, more extended, portions of the jet. We also included VLBA data at 15 (for visualization) and 24 GHz.

We fit a single-zone leptonic model to the SED,  using the Jets SED modeler and fitting Tool (\texttt{JetSeT})\footnote{\url{https://jetset.readthedocs.io/en/1.1.2/}} \citep{Tramacere2009,Tramacere2011,jetset_ascl}
for a 
Synchrotron Self Compton (SSC) + External Compton (EC) scenario. The accretion disk spectrum is modeled as multi-temperature blackbody as described in \cite{Accretion_Power_2002}, with a luminosity $L_{\rm Disk}$, a black hole mass  $M_{BH}$ (fixed to the value derived in \citealt{lore2018}), and an accretion efficiency fixed to $0.08$. The BLR is assumed to be a clumpy thin spherical shell with and internal radius determined by the phenomenological relation provided by \cite{Kaspi_2007},  $R_{BLR,in}=3\times10^{17}L_{\rm Disk,46}^{1/2}\,$cm. The external radius of the BLR is assumed to be $0.1 R_{\rm BLR,in}$, with a coverage factor $\tau_{BLR}=0.1$. The dusty torus (DT) radiation is assumed to be described by spherical uniform radiative filed, with a radius $R_{DT}=2\times10^{19}L_{\rm Disk,46}^{1/2}\,$cm,  \citep{Cleary_2007}, and a reprocessing factor $\tau_{DT}=0.1$. Both the radius of the DT, and the radii of the BLR are implemented in \texttt{JetSeT} as dependent parameters, hence during the fit they are not free but determined by the phenomenological relation described above. The emitting region is assumed to be a single zone with a spherical geometry and radius $R$, located at a distance $R_{H}$ from the central black hole of mass logM$_{BH}$ = 8.4. The jet is assumed to be conical at the scale where the emitting region is located, with an half opening angle of $\phi\approx$ 5 deg, with the emitting region size determined by  $R=R_{H} \tan{\phi} $. The blob moves through the jet with a bulk Lorentz factor $\Gamma$, oriented at a viewing angle $\theta$, and a consequent beaming factor $\delta= 1/(\Gamma \sqrt{1-\beta_{\Gamma} \cos(\theta))}$. The relativistic electrons are assumed to follow a broken power-law energy distribution,
 \begin{equation}
     n(\gamma)= N \left\lbrace
     \begin{array}{ll}
     N_0\gamma^{-p} & \gamma_{min}\leq \gamma \leq \gamma_{b} \\
    N_0\gamma^{-p_1}\gamma_{b}^{p-p_1} & \gamma_{b}<\gamma<\gamma_{max},
     \end{array}
     \right.
 \end{equation}
with an index of $p$ and $p_1$ below and above the break energy $\gamma_b$, respectively. The electron distribution normalization constant, $N_0$, is set in order to have $\int^{\gamma_{max}}_{\gamma_{mind}} n(\gamma) d\gamma=N$. Given the small size of the BLR we set 
the initial position of the emitting region at $R_H\approx 1$ pc. The
resulting best fit model is shown in Figure \ref{fig:sed}. The initial value of $L_{\rm Disk}$, is determined by \texttt{JetSeT} during the pre-fit stage, and is set to an initial value of   $L_{\rm Disk}=10^{43}$ erg s$^{-1}$.   The minmization of the model is performed using the \texttt{JetSeT} \texttt{ModelMinimizer} module plugged to \texttt{iminuit} python interface \citep{iminuit}. The errors are estimated from the matrix of second derivatives, using the \texttt{HESSE} method. Even though, in \texttt{JetSeT} errors can be evaluated more accurately forcing the \texttt{MINOS} \texttt{iminuit} method, or by using a MonteCarlo Markov Chain, for the current analysis the \texttt{HESSE} method provides a fair estimate. We fit data above 20 GHz, excluding data below the synchrotron self-absorption frequency, we also
add a 10\% systematic error to data below $10^{16}$ Hz to avoid that the small errors in the UV-to-radio frequencies biases the fit toward the lower frequencies.
Since  the model fit returned for both the sates (2015 and 2018)  a similar best-fit  value of $L_{\rm Disk}\approx10^{43}$ erg s$^{-1}$, and of $\theta \approx 3$ deg, then  we decided to freeze these parameters.

\begin{table}
\caption{Parameters of the fitting for the SED built with the 2015 data, and the SED built with the 2018-19 data. The parameters marked with * were frozen in the fit, the parameters marked with $^{\dagger}$ are dependent parameters (see text for details).}
\label{tab:jetset_fit_new}

\begin{center}
\begin{tabular}{llll}

\hline
Name              & Units      & Values (2015)                        & Values (2018-19)                 \\
\hline
$\gamma_{min}$*    &            & 1                & 1                            \\
$\gamma_{max}$    &            & $(4.0 \pm 0.2)\times 10^3$ & $(4 \pm 3)\times 10^3$                       \\
$N$               & cm$^{-3}$  & $(2.4 \pm 0.8)\times 10^3$  &  $(8 \pm 2)\times 10^2$                        \\
$\gamma_{b}$      &            & $1917 \pm 221$                & $110 \pm 10$                                     \\
$p$               &            & $2.3 \pm 0.2$              & $2.2 \pm 0.8$                                    \\
$p_1$             &            & $3.4 \pm 0.4$              & $3.7 \pm 0.2$                                      \\
$T_{\rm DT}$*      & K          & $330$                        & $330$                                            \\
$R_{\rm DT}$$^{\dagger} $     & cm         & $6.3\times 10^{17}$          & $6.3\times 10^{17}$                        \\
$\tau_{\rm DT}$*   &            & $0.1$                        & $0.1$                                             \\
accr. eff.*        &            & $0.08$                       & $0.08$                                            \\
$M_{\rm BH}$*      & $M_{\sun}$ & $2.5\times 10^8$             & $2.5\times 10^8$                                  \\
$\tau_{\rm BLR}$*   &            & $0.1$                        & $0.1$                                             \\
$R_{\rm BLR,in}$$^{\dagger} $  & cm         & $9.5\times 10^{15}$            & $9.5\times 10^{15}$                        \\
$R_{\rm BLR,out}$$^{\dagger} $ & cm         & $1.0\times 10^{16}$            & $1.0\times 10^{16}$                        \\
$L_{\rm Disk}$    & erg/s        & $10^{43} $             & $10^{43}$                                  \\
$R$$^{\dagger}$               & cm         & $4.8 \times 10^{16}$  & $6.3\times 10^{16}$                    \\
$R_{H}$           & cm         & $(4.76 \pm 0.09) \times10^{17}$    & $(6.325\pm 0.006) \times10^{17}$                           \\
$B$               & G          & $0.20 \pm 0.01$           & $0.42 \pm 0.05$                                     \\
$\theta$*          & deg        & $3.0$                       & $3.0$                              \\
$\Gamma$          &            & $24 \pm 2$                 & $19 \pm 5$                                       \\
\hline
\end{tabular}
\end{center}
\end{table}

We used the same model to fit both, the data in 2018, and the data in the SED presented in \cite[][using data from 2015]{lore2017} using data from the VLBA in the radio, \emph{XMM-Newton} in UV/optical/X-rays, and added the gamma-ray upper limit obtained for Fermi-LAT data within the dates mentioned in Section \ref{gammarays}. 
We also included data from the 4FGL-DR3 in three energy bands (0.3-1, 1-3, and 3-10 MeV) in comparison with the upper limits in both SEDs. However these data are for visualization purposes only because flares can easily exceed stacked average constraints and we do not know what variability behavior might have gone into the stacked average including 12 years of data.

Both the 2015 and 2018 SEDs can be explained by an EC dominated scenario coming from the dusty torus, with mild contribution of the SSC component.
The most relevant model parameters are reported in Table \ref{tab:jetset_fit_new},
and the best fit SEDs are shown in Figure \ref{fig:sed}.  
 Due to the lack of data in the mm-IR band, where the peak of the synchrotron component occurs, it is not straight forward to provide a firm estimate of the luminosity of this component from the model fit, anyhow we notice that the total energetic budget of the jet has not showed a dramatic change. Assuming a barionic load of one cold proton each lepton, the total luminosity of the jet in 2015 is of $\approx 4\times 10^{47}$ erg s$^{-1}$, and of $\approx 2\times 10^{47}$ erg s$^{-1}$  in 2018, with a radiative power changing   from  $\approx 6\times 10^{42}$ erg s$^{-1}$ in 2015, to 
 $\approx 2\times 10^{42}$ erg s$^{-1}$ in 2018. 
 The difference in the radiative output can be explained mainly by the changes in the bulk Lorentz factor, and in the electron distribution, with the one in 2015 being harder and characterized by larger value of $\gamma_{b}\approx 1900$. Anyhow, a model with spectral changed due mostly to variation of  $\Gamma$ and/or different viewing angles and magnetic field intensity, could  still be accommodated. In conclusion, we think that the behaviour of the jet, based on the result of the SED modeling, can be related to a mostly quiescent state of the source, with a moderate change in the radiative output and moderate flarign activity in a shock located at a scale between 0.1 and 1 pc.

\section{Discussion}

\subsection{Variability of AGN and blazar populations}

Variability is a property characterizing AGN, that manifests in a variety of timescales, ranging from minutes to years depending on the type of source \citep{netzer2013}. Among them, some blazars show the most peculiar variability pattern, showing flaring activity with high amplitude variations (in some cases to some orders of magnitude) in timescales as short as a few days, as shown by intensive monitoring campaigns of particular sources \citep[e.g.,][]{patino2018, magic2018, fraija2019, fernandes2020, vahram2020, zargaryan2022, guise2022, priya2022, magic2022}. However, intensive monitoring is usually performed for interesting sources that usually show the largest amplitude variations (in particular for blazars), so these studies could be hampered by the few available resources to follow-up large samples of sources at different wavebands.

The ZTF survey has a sampling rate of three days with the possibility to monitor light curves of large samples of AGN, especially blazars. This allows a comparison of the optical variability properties of these samples with the ones of \pbc.

The samples used for comparison were taken from the training set used by the ALeRCE light curve classifier \citep[see][for details]{paula2021}. The first sample is composed by a total of 4667 non-blazar AGN; these sources were taken from the class ``A'' of the Million Quasars Catalog (MILLIQUAS, version 6.4c, 2019 December; \citealt{flesch2015,flesch2019}) and the New Catalog of Type 1 AGNs \citep{oh2015}. The second sample contains 1267 blazars, these were taken from the 5th Roma-BZCAT Multi-Frequency Catalog of Blazars (ROMABZCAT; \citealt{massaro2015}), and the class ``B'' sources of the MILLIQUAS\footnote{Class ``A'' are type-I Seyferts/host-dominated AGN, and class ``B'' are BL Lac type object in MILLIQUAS.}.

\begin{figure}
	\includegraphics[width=9cm]{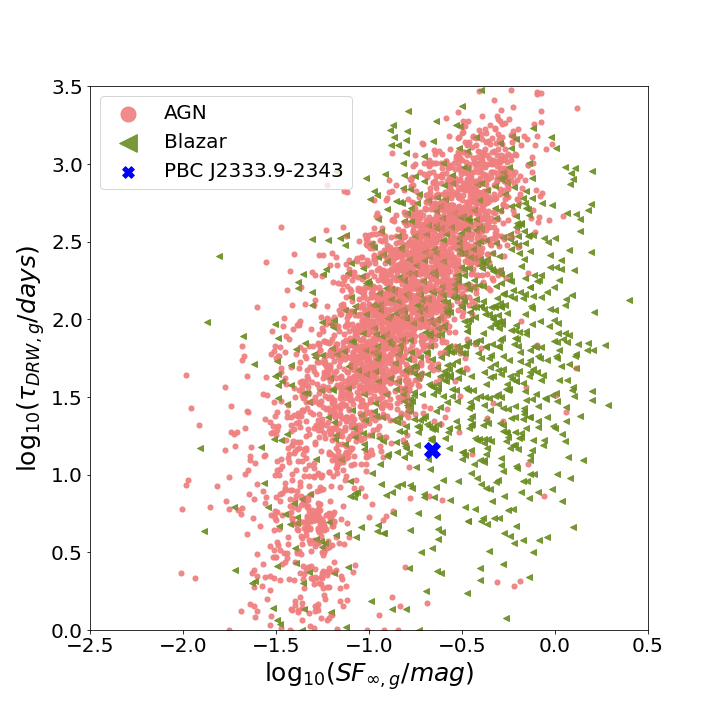}
\caption{Distribution of the Damp Random Walk (DRW) parameters, $\tau_{DRW}$ (in days) against $SF_{\infty}$ (in $mag$), in logarithmic scale, in the g-filter for the ALeRCE training set sample of non-blazar AGN (red circles) and blazars (green triangles). The parameters for PBC\,J2333.9-2343 are marked as a blue cross. }
    \label{fig:drw}
\end{figure}

In particular, we used the ALeRCE light curve classifier repository\footnote{\url{https://github.com/alercebroker/lc\_classifier}} to compute variability features.
\cite{ruan2012} have shown that the Damped Random Walk \citep[DRW,][]{kelly2009} parameters are able to differentiate the variability properties of blazar and non-blazar populations of AGN.
We used the ZTF alerts light curves for this comparison, as well as the complete alert light curve of PBC\,J2333.9-2343\footnote{As a reminder, we presented the forced photometry light curve during the monitoring, but for this comparison with a large sample of sources, we use the alerts light curve (see Section \ref{ztf}).} retrieved from ALeRCE. We measured $\tau_{DRW}$; the characteristic time for the time series to become roughly uncorrelated; and $\sigma_{DRW}^2$; the squared amplitude of the variations; for the g-filter (that is the less affected by the host galaxy contribution) of the ZTF light curves. 
From $\sigma_{DRW}^2$ we estimated the asymptotic value of the structure function on long timescales as SF$_\infty = \sqrt{2} \sigma_{DRW}$ \citep{macleod2011}.
The measurements are shown in Fig. \ref{fig:drw}. AGN are represented as red circles, blazars as green triangles, and PBC\,J2333.9-2343 is marked with a blue cross, and is among the blazar population, suggesting its optical variability properties are closer to those of the blazar population. This plot shows results in agreement with \cite{ruan2012}, who reported that blazars have $\tau_{DRW}$ in between those of normal quasars and other objects (mainly variable stars), as well as larger values of SF$_\infty$.
It is worth remarking that there might exist a degree of misclassification between blazar and non-blazar AGN. Indeed, the confusion matrix in \cite{paula2021} shows that 74$^{+5}_{-3}$\% of blazars are well classified using the ALeRCE light curve classifier, while the rest of the sources are classified as AGN or Quasar. For instance, blazars could be classified as AGN during non-flaring activity, explaining why there is some mix between the classes.

\subsection{Constraints on the region responsible for the variations observed in the optical/NIR bands}

 Optical and NIR variations are normally seen in radio quiet AGN as well as radio loud objects. In the former case they are often attributed to variations in the accretion disc (optical and NIR) and reprocessing of variable emission in the torus (NIR). In this section we compare the upper limits on the delay between I band and K band fluctuations in \pbc\ to the ranges of delays expected in these non-jet scenarios, in order to constrain the origin of the variable optical/NIR emission in this object.  
In \pbc\ we found a time lag between the K and I band of 1.02$\pm$1.45 days, so we can investigate if the lag would be compatible with the K band emission arising from the torus, the accretion disk or the jet. We use a black hole mass of logM$_{BH}$ = 8.4 and the luminosity at 5100 \AA\ reported in \cite{lore2018} to estimate an Eddington ratio $R_{\rm Edd}$ of 3.2$\times$10$^{-3}$ following \cite{paula2018}.

When the NIR emission is dominated by the torus,
the dust, responsible for the re-emission in the NIR, can only be located at distances of at least light days/weeks, so variations may occur smoothly and with time lags of weeks/months due to the different time delays on different sides of the torus \citep{lira2015,paula2017}. Indeed, recent studies show delayed response of the K-band light curve after the V-band light curve of at least 10 days, and in particular for the luminosity of \pbc\ a lag longer than 80 days is expected \citep{koshida2014, minezaki2019}. We can therefore reject that the torus is responsible for the NIR emission.

The variable optical and NIR emission can also come from the accretion disk, either through intrinsic (i.e. viscosity or thermal) changes in the disk or through reprocessing of a variable illuminating source, for example of X-rays \citep{netzer2013}. In the first case, the intrinsic variations have difficulty reproducing quasi-simultaneous flux changes in two different bands, unless it is only a small region of the disk which is varying and this region contributes all the \emph{variability} of both bands even if it does not produce all their \emph{emission}. In any accretion disk where the surface temperature decreases outwards, longer wavelength are emitted by larger fractions of the disk. Therefore, if only a small central region of the disk is producing the variations, these will modulate a larger fraction of the total I band emission than of the total emission of the disk in the K band and the fractional variability of the I band would be larger than that of the K band, which is contrary to what we observe.
On the other hand, to estimate the delay expected from reprocessing on the disk, we followed \cite{lira2015} to estimate the light travel time across a standard accretion disk model as $\tau = 3 \times 10^{-10}\lambda^{4/3}R_{Edd}^{1/3}M_{BH}^{2/3}$, with $\lambda$ in \AA, $\tau$ in days, $R_{Edd}$ in Eddington units and M$_{BH}$ in solar masses. The expected lag between the K and I band is of 7.5 days, much larger than the measured value of $1.02\pm{1.45}$ days. Thus it is unlikely that the accretion disk is responsible for the variable emission in both I and K bands.

If variations are related to jet emission and the jet is oriented towards the observer, however, the variability timescale is shortened and the radiation is strongly enhanced by relativistic beaming. The characteristic timescales of such variations are of about a few days, which may be interpreted as the typical timescale of successive flare events \citep{kataoka2001}. In fact, if the emission is dominated by emission from the jet both at optical and NIR frequencies, then the variations are rapid, high amplitude and simultaneous in both bands \citep{asmus2015}. For instance, \cite{bonning2012} presented the results of light curves from a sample of 12 blazars observed by SMARTS-1.3m in the BVRJK between 2008-10, showing that their CCFs are centered around zero.
The monitoring campaign presented here, as well as the analysis of the variability in PBC\,J2333.9-2343 (see Sect. \ref{variability}), reveals large amplitude variations at short timescales showing flaring behaviour, and the fact that the variations occur simultaneously in the optical and NIR band agrees well with the blazar nature of PBC\,J2333.9-2343.

The doubling/halving times were estimated to be 7.7 days (I band) and 6.7 days (K band). We estimated the electrons cooling time in the observer frame using best fit parameters for the SED from 2018-19, taking into account both synchrotron and IC/EC in Thomson regime, which results to be $\sim$0.3 days, and the escape time, ignoring possible contribution from turbulence, ($\frac{R(1+z)}{c \delta_D }$) $\sim$ 1 day. 
This seems to indicate that the decay could be driven not only by the electrons cooling/escape times, but other mechanisms may play a role, as for instance a modulation in the injection of the particles (and/or in the beaming pattern), or an area of magnetic reconnection in the flare \citep[see e.g.,][]{zhu2018, sahakyan2020, pandey2022}.

\subsection{Multiwavelength emission}

Extra-galactic emission in the gamma-ray band is a phenomenon related to the presence of relativistic jets \citep{dermer2016}. PBC\,J2333.9-2343 is detected by Fermi-LAT and reported in the fourth catalog of AGN detected by the Fermi Gamma-ray Space Telescope Large Area Telescope (4LAC), with an integrated flux of 3.9$\pm$0.5$\times$10$^{-12}$ erg$\hspace{0.1cm}$s$^{-1}$cm$^{-2}$ in the 0.1-100 GeV band between 2008 and 2016, and a spectral index of 2.42$\pm$0.12 \citep{ajello2020}, in agreement with our results (see Sect. \ref{gammarays}). Indeed, this source is classified as a remarkable `other AGN' case because it was not possible to classify it. 
We notice that the fact that gamma-ray emission was detected on this source with Fermi-LAT at a 6.2$\sigma$ of significance level is more likely indicative to be from a blazar rather than a radio galaxy. Furthermore, our results show that this source has varied in the gamma-ray range, the number of photons being almost half in 2015 compared to the 2018-2019 data. This is in agreement with the variability results from the 4FGL-DR3 where a variability index of 30.8\footnote{A value greater
than 24.72 over 12 intervals indicates $>$99\% chance to be a variable source.} is reported for this source \citep{fermidr3}.

Blazars characteristically show a two hump structure (low and high energy), as we see for PBC\,J2333.9-2343. The low energy hump is well explained by synchrotron emission by ultrarelativistic electrons, whereas the high peak is usually interpreted as inverse Compton (IC) emission for the case of a pure leptonic scenario \citep{blandford1979}, or dominated by high energy emission of ultra-relativistic protons in the case of the hadronic scenario \citep{dermer1993,bottcher2013}. Radio galaxies can also show a two hump structure, the way to differentiate between a blazar or a radio galaxy being the jet angle \citep[e.g.,][]{ghisellini2005}.
In this work we used a one-zone leptonic model, where the high energy frequencies are well fitted by EC emission and some SSC contribution, with values of the model parameters within the blazar expectations, among them a jet angle of 3 degrees (see Fig. \ref{fig:sed} and Table \ref{tab:jetset_fit_new}).

A previous SED fitting was presented in \cite{lore2017} using simultaneous \emph{XMM--Newton} data. The main difference between the SED fittings is that in \cite{lore2017} the SSC and EC are fitted separately, whereas here they are fitted simultaneously. Moreover, 
the photon contribution to the EC model in \cite{lore2017} comes from IC emission with seed photons from the torus, whereas in this work the contribution comes from IC emission with seed photons from the disk, the BLR and the torus.

In order to be able to compare the SEDs built in 2015 and 2018, we fitted the two of them with the same model (see Section \ref{SEDsect}).
The main result from this analysis is that the jet angle must be very close to the line of sight of the observer.
For larger angles the Doppler boosting is smaller and therefore in order to properly fit the lower frequency peak of the SED it becomes necessary to  either increase the energy of the electron population, or increase the amount of electrons available (either by increasing the density or the size of the region). However, for both cases, the high energy part of the SED model increases significantly, which results in a poor fit to the X-ray data. Therefore, we found necessary to increase the Doppler boosting (lowering the angle) to obtain a good fit.

The advantage of the current SED is that more data points are available, including data in the gamma-rays, which allows a more robust fit. 
In particular, the fact that gamma-ray emission is detected and the SED shows a two peaked structure favors the blazar-like nature hypothesis, consistent with its variability pattern. It is worth remarking that according to the SED modeling, in this object, the contribution of the disk and the torus is negligible compared to the synchrotron emission, which would be responsible for almost all the optical and NIR emission, as well as the X-rays and gamma-rays can  also be explained by emission reprocessed by the jet, with mild contribution from the dusty torus.

Blazars are further classified into flat-spectrum radio quasars (FSRQ) and BL Lac (BL Lacertae being the prototype) based on various observational properties. BL Lacs show no (or weak) emission lines in their optical spectra, and the synchrotron peaks in the SED at frequencies $>$ 10$^{14}$ Hz, whereas FSRQ do show emission lines and their synchrotron typically peaks at frequencies lower than 10$^{14}$ Hz \citep{abdo2010,giommi2012,padovani2017}. The optical spectrum of PBC\,J2333.9-2343 shows prominent narrow and broad emission lines \citep{lore2018} and the synchrotron peak in the SED is at $<$ 10$^{14}$ Hz (Fig. \ref{fig:sed}), therefore we classify the nucleus of this galaxy as a FSRQ.

We would like to stress the peculiarities of this galaxy, with emphasis on the low gamma-ray luminosity ((3.10$\pm$0.02)$\times$10$^{43}$ erg $s^{-1}$) compared to other blazars \citep[see Fig. 10 in][]{ajello2020}, more typical of BL Lacs.
It has also being suggested that BL Lacs are dominated by SSC whereas FSRQ are dominated by EC, in agreement with the classical FSRQ classification of PBC\,J2333.9-2343. However, there are examples in the literature of sources that can change from EC to SSC dominated, as for example 3C\,279 \citep{patino2018}.
In the recent work by \cite{pei2022} the authors proposed an ``appareling zone'' consisting on a potential transition field between BL Lacs and FSRQs where changing-look blazars may reside based on four physical parameters, where \pbc\ indeed should be located. Other examples of sources showing observational characteristics changing between a BL Lac and a FSRQ include BS\,B1646+499 \citep{pajdosz2018} or B2\,1420+32 \citep{mishra2021}. Observing morphologically different AGN types simultaneously has been proposed as the result of jet axis reorientation \citep{Pajdosz2022}. PBC\,J2333.9-2343 could also be an extreme case of XRG, with the new jet pointing towards us and therefore preventing us from observing the X-shaped morphology.


\begin{figure}
	\includegraphics[width=10cm]{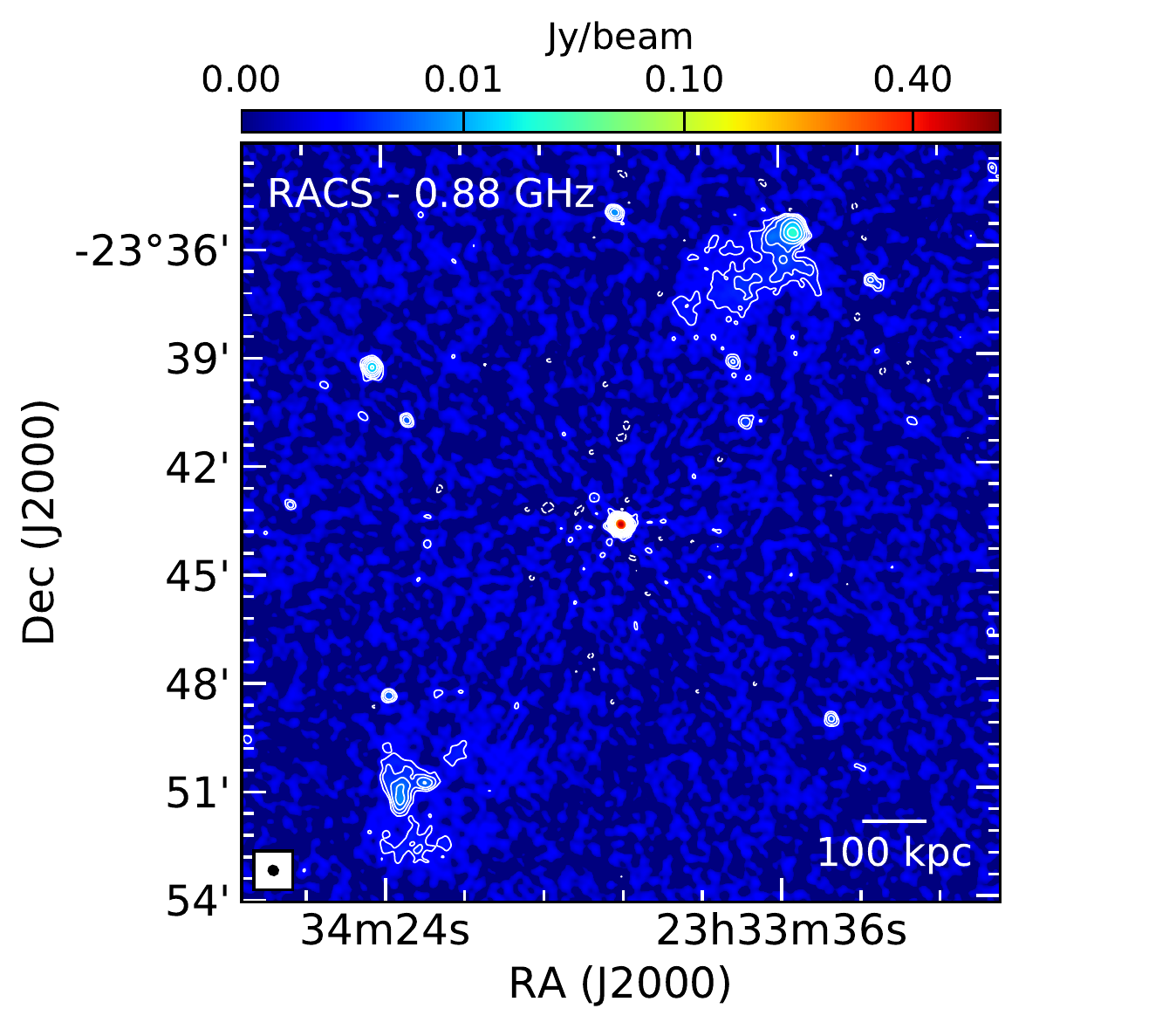}
\caption{Image of PBC\,J2333.9-2343 from the RACS survey at 0.88 GHz. Contours are 3$\times$RMS$\times$(-1, 1, 2, 4, 8, 16, 32, 64, 128, 256). The HPBW (14.8\arcsec$\times$13.5\arcsec) is shown in the lower-left corner.}
    \label{fig:racs}
\end{figure}


\subsection{\label{radiodisc}Radio morphology and long-term variability from the latest radio surveys and archives}

An additional evidence in favour of the blazar nature of the core is the lack of clear jet emission connecting the lobes and the core region presented in \cite{bruni2020} with the GMRT at 150 MHz. To further confirm this result, we considered images of the source from recently released radio surveys: the VLASS at 3 GHz, and the RACS at 0.88 GHz. The latter has a sensitivity ten times larger than the mentioned GMRT observations, and five times larger than the NVSS. Moreover, the short baselines of ASKAP allow to recover extended structures up to 1 deg, making it suitable for the study of GRGs. The RACS image is presented in Figure \ref{fig:racs}: at a noise level of 215 $\mu$Jy/beam, and a resolution of $\sim$15\arcsec, no sign of connection between the lobes and the nucleus is visible, confirming previous results. The shortest distance between the core and the first contour of the lobes is $\sim$3\arcmin~(167 kpc at the redshift of the source), resulting in a missing association and the consequent absence of this source in the recent catalogue of GRG from RACS \citep{2021Galax...9...99A}.
The VLASS quick-look image, at an angular resolution of $\sim$2.8 arcsec and an RMS of 580 $\mu$Jy/beam, detects only the core. This is expected, since the VLA configuration used for the survey (B) only allows to recover structures with an angular size up to 1\arcmin. However, the absence of a jet even in the regions closer to the core, confirms the discontinuity visible at lower frequencies.

Finally, we have investigated the core long-term radio variability browsing the NRAO VLA archive survey (NVAS\footnote{\url{http://www.vla.nrao.edu/astro/nvas/}}, \citealt{2007AAS...21113203C}). We could collect images between 1983 and 2001, covering almost 20 years at 4.8 and 15 GHz. At 8.5 GHz, the observations span about 10 years (1990-1999). The highest flux density was recorded at 15 GHz during November 1984 (6.9$\pm$0.3 Jy), corresponding to a factor of $\sim$ 7 of variability with respect to previous and subsequent epochs. A corresponding, although lower, peak is present at 4.8 GHz (200\% increase, also November 1984), confirming the variation. Only mild variations were recorded at these frequencies during later years. At 8.5 GHz, the absolute maximum was recorded in July 1991 (an increase of a factor 2 with respect to the previous epoch), and milder variations during 1999. As a whole, these archival data confirm the pronounced variability discussed in this work, and detected in the radio band with the Effelsberg observations on a shorter time window, as well as comparison with previous works such as the flux reported at 8 GHz by \cite{ojha2004} and \cite{lore2017} with data obtained 11 years later.


\subsection{A re-oriented jet or an intervening blazar in a GRG?}

Another possibility to explain the behaviour of this source is to consider the presence of an intervening blazar in our line of sight. 
Looking at the number of hard X-ray selected blazars, N, as a function of flux, S (14-195 keV), logN-logS (see figure 10 from \citealt{Langejahn2020}) and assuming the \emph{Swift}/XRT error circle of 6 arcsec, the number of blazars similar to \pbc\ expected to fall in it is $\sim$6$\times$10$^{-5}$, i.e. indicating that a chance overlap is unlikely.

Generally in these cases there is an overlapping of two different optical spectra with some peculiar composition, as for instance lines at different redshift, that should appear as double peaked lines in the narrow emission lines. This is not observed in the spectra of \pbc\  even with data obtained from the Very Large Telescope at a dispersion of 0.3 \AA/pixel (Hernandez-Garcia et al., in prep.). 
Then, if there are two sources we should assume that there is a broad-line radio galaxy with its optical spectrum and then a blazar with a featureless blue spectrum which becomes embedded in the radio galaxy. However, no indications of composite spectral continuum are found.

If an intervening blazar can be discarded, we can conclude that the GRG and the blazar nucleus in \pbc\ are part of the same galaxy. Then, the most plausible explanation is that the jet has changed its direction, as previously proposed in \cite{lore2017}, making this an exceptional case of jet reorientation. These kind of changes have already been proposed to explain XRGs.
For example, \cite{dennett2002} studied two XRG (3C\,223.1 and 3C\,403) and explained the morphology in terms of a rapid realignment of the radio jet, and considered a binary black hole merger or acquisition of a smaller galaxy as likely candidates for the cause of the change of the jet axis, as they did not see any indication of merging. 
Similarly, \cite{machalski2016} reported the case of 3C\,293, which is very likely the result of a post merging event with the galaxy UGC\,8782. In this work they concluded that the jet axis flipped rapidly due to tidal interaction of its merging process.
In the particular case of PBC\,J2333.9-2343, we do not see the X-shape, but this can be explained because the new jets are, by chance, pointing towards us. The confirmation of no emission between the nucleus and the jet discussed in Sect. \ref{radiodisc} strongly supports the idea of a re-oriented jet.

\section{Conclusions}

In this work we presented a contemporaneous multiwavelength monitoring of the nucleus in \pbc\ that covered two periods, between September 2018 and January 2019, and April-July 2019.
Variations are found at all observed wavelengths at significance larger than 6$\sigma$, except at X-rays where variations are detected at 2$\sigma$ of confidence level within the four month monitoring period. The observed variations occur in timescales shorter than a month and with amplitudes larger than a factor of two. The cross-correlation between optical/NIR also shows that the variations occur simultaneously in these bands. When comparing the optical variability features with large samples of non-blazar AGN and blazars, PBC\,J2333.9-2343 shows characteristics more similar to the blazar population. According to these results, we interpret the observed variations as flaring events.

We constructed the SED, that we then fitted using a single-zone leptonic model. The SED shows two distinct peaks, the low energy one is well fitted by synchrotron emission, while the high energy peak is dominated by EC from the torus with some contribution from SSC. This SED was compared with the data already presented in \cite{lore2017} using VLBA and \emph{XMM-Newton}, and we added Fermi-LAT data. 
The jet angle in the fitted models is 3 degrees, indicative of a blazar.

These results and the gamma-ray detection at 6$\sigma$ of confidence level strongly suggest the presence of a blazar-like nucleus at the center of \pbc . This galaxy was previously classified as a GRG, suggestive of a change in the direction of the jet as previously proposed in \cite{lore2017}. Further evidence in agreement with this scenario is the fact that no connection between the nucleus and the lobes is observed in the deepest radio images, and historical radio fluxes from the NRAO VLA archive survey revealed variations by a factor of seven about 30 years ago, confirming the pronounced variability in the radio at 20.4 GHz shown in the present work.

In the future we can use resources such as the ZTF or ATLAS alert streams to monitor this or other interesting sources and trigger other instrumentation at different wavebands when a flaring event is detected. In particular, ALeRCE has a watchlist service\footnote{\url{https://watchlist.alerce.online/}} that notifies via email when a source from your own target list generates an alert. 

\section*{Acknowledgements}

We acknowledge funding from ANID programs:  Millennium Science Initiative ICN12\_009 (LHG, AMA, PSS, FEB, FF) and NCN$19\_058$ (PA, PL); CATA-Basal - ACE210002 (FEB), FB210003 (FEB, FF) and  BASAL project FB210005 (AMA); and FONDECYT Regular 1190818 (FEB) and 1200495 (FEB).
G.B. and F.P. acknowledges financial support under the INTEGRAL ASI-INAF agreement 2019-35-HH.0 and ASI/INAF n. 2017-14-H.0.
This work was partially supported by CONACyT (Consejo Nacional de Ciencia y Tecnología) research grants 280789 (V.M.P.-A., VC, México) and 320987 (V.M.P.-A.,México). 
This work is supported by the MPIfR-Mexico Max Planck Partner Group led by V.M.P.-A, and the MPA-Universidad de Valparaíso Max Planck Partner Group led by PA. 
VC acknowledges support from the Fulbright - García Robles scholarship.
This publication has received funding from the European Union's Horizon 2020 research and innovation program under grant agreement No. 730562 (RadioNet).
This research has used data from the SMARTS-1.3m telescope, which is operated as part of the SMARTS Consortium, as part of the approved proposals CN2018B-19,and CN2019A-Fast Track (CNTAC).
We acknowledge the use of public data from the \emph{Swift} data archive (ToO ID 10939).
Partly based on observations with the 100-m telescope of the MPIfR (Max-Planck-Institut für Radioastronomie) at Effelsberg.
We thank the staff of the Effelsberg-100m telescope, for making these observations possible (Proposal 15-18). 
The ZTF forced-photometry service was funded under the Heising-Simons Foundation grant
\#12540303 (PI: Graham).
This work has made use of data from the Asteroid Terrestrial-impact Last Alert System (ATLAS) project. The Asteroid Terrestrial-impact Last Alert System (ATLAS) project is primarily funded to search for near earth asteroids through NASA grants NN12AR55G, 80NSSC18K0284, and 80NSSC18K1575; byproducts of the NEO search include images and catalogs from the survey area. This work was partially funded by Kepler/K2 grant J1944/80NSSC19K0112 and HST GO-15889, and STFC grants ST/T000198/1 and ST/S006109/1. The ATLAS science products have been made possible through the contributions of the University of Hawaii Institute for Astronomy, the Queen’s University Belfast, the Space Telescope Science Institute, the South African Astronomical Observatory, and The Millennium Institute of Astrophysics (MAS), Chile.
The ASKAP radio telescope is part of the Australia Telescope National Facility which is managed by Australia’s national science agency, CSIRO. Operation of ASKAP is funded by the Australian Government with support from the National Collaborative Research Infrastructure Strategy. ASKAP uses the resources of the Pawsey Supercomputing Research Centre. Establishment of ASKAP, the Murchison Radio-astronomy Observatory and the Pawsey Supercomputing Research Centre are initiatives of the Australian Government, with support from the Government of Western Australia and the Science and Industry Endowment Fund. We acknowledge the Wajarri Yamatji people as the traditional owners of the Observatory site. This paper includes archived data obtained through the CSIRO ASKAP Science Data Archive, CASDA (https://data.csiro.au).
The National Radio Astronomy Observatory is a facility of the National Science Foundation operated under cooperative agreement by Associated Universities, Inc. 
CIRADA is funded by a grant from the Canada Foundation for Innovation 2017 Innovation Fund (Project 35999), as well as by the Provinces of Ontario, British Columbia, Alberta, Manitoba and Quebec. 
We thank the ALeRCE Broker for making their services public to the scientific community. In this work we used their Web Interface and the ZTF Forced Photometry Notebook.
%


\section*{Data availability}

The data underlying this article were accessed from the ZTF Forced Photometry Service (https://ztfweb.ipac.caltech.edu/cgi-bin/requestForcedPhotometry.cgi), the ATLAS Forced Photometry Service (https://fallingstar-data.com/forcedphot/), \emph{Swift} (https://swift.gsfc.nasa.gov/), and Fermi (https://fermi.gsfc.nasa.gov/) archives.
The Effelsberg, VLBA, SMARTS-1.3m data and the derived data generated in this research will be shared on reasonable request to the corresponding author.



\bibliographystyle{mnras}
\bibliography{mnras_template} 

\newpage


\appendix

\section{\label{appendix}Figures and tables}

\begin{figure}
	\includegraphics[width=10cm]{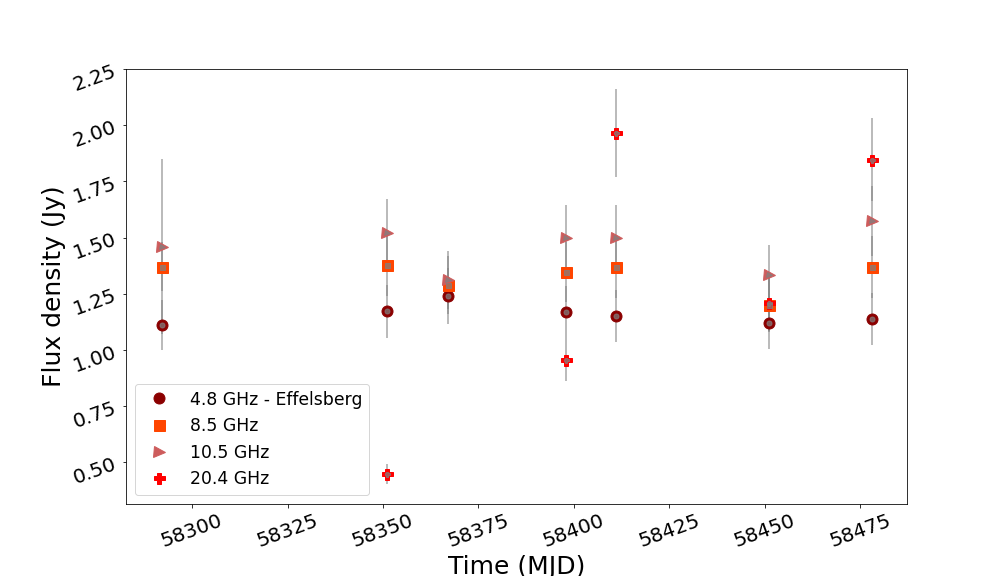}
\caption{Effelsberg radio flux densities collected from June to December 2018, at 4.8, 8.5, 10.5, and 20.4 GHz.}
    \label{fig:Eff}
\end{figure}

\begin{figure}
	\includegraphics[width=10.0cm]{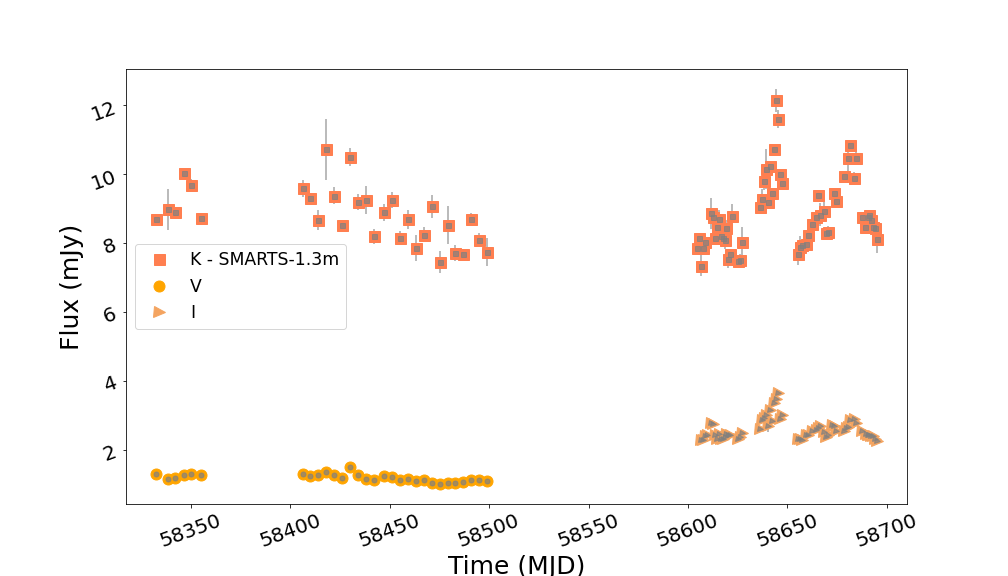}
\caption{Flux light curves in $V$ and $I$ bands and the corresponding simultaneous $K$ band light curves with SMARTS-1.3m.  }
    \label{fig:smarts}
\end{figure}

\begin{figure}
	\includegraphics[width=10.0cm, angle=0]{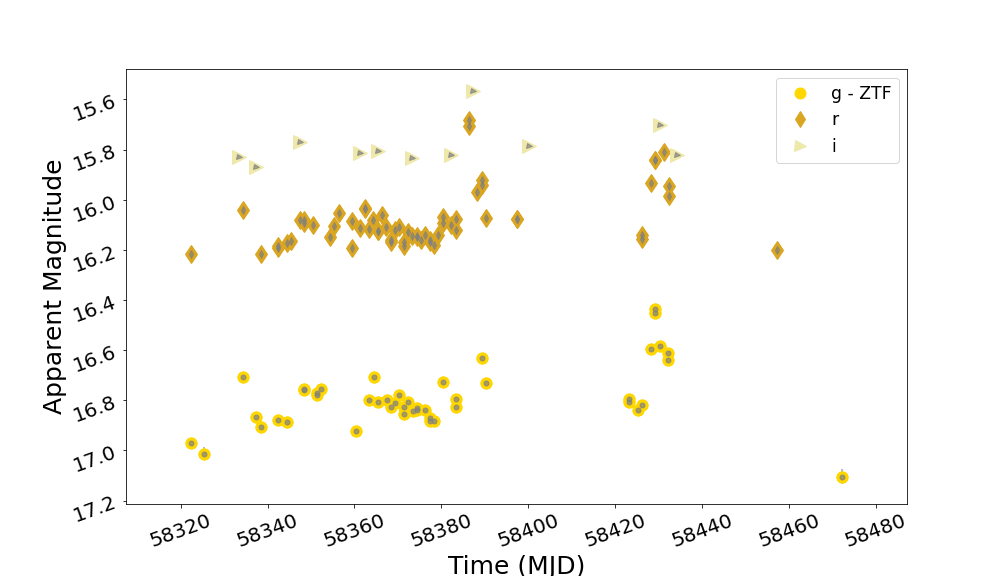}
\caption{\label{fig:ZTF}ZTF forced photometry light curves in the gri filters between July 2018 and January 2019. }
\end{figure}

\begin{figure}
	\includegraphics[width=9.0cm, angle=0]{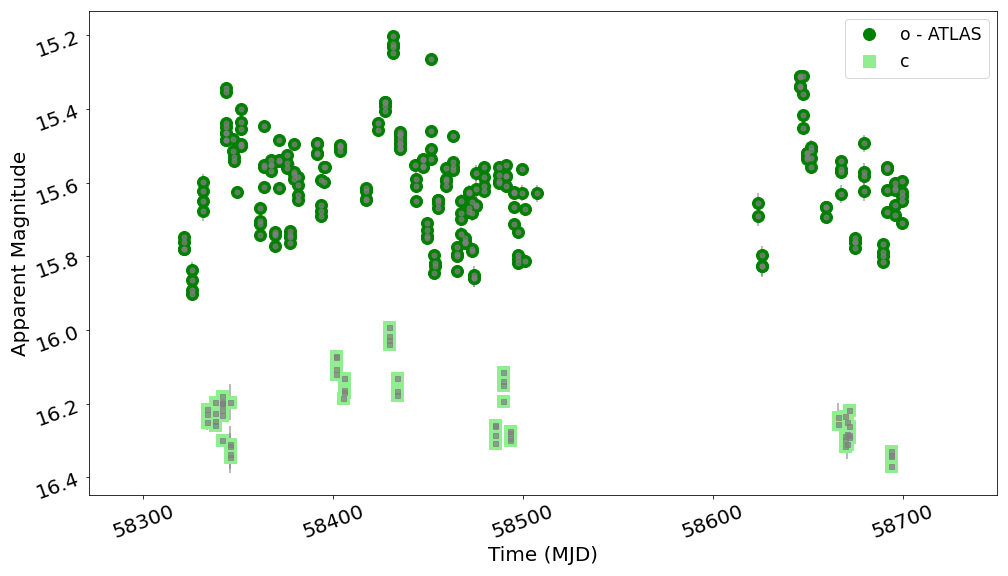}
\caption{\label{fig:ATLAS}ATLAS forced photometry light curves in the c and o filters between July 2018 and May 2019. }
\end{figure}

\begin{figure}
	\includegraphics[width=10.0cm]{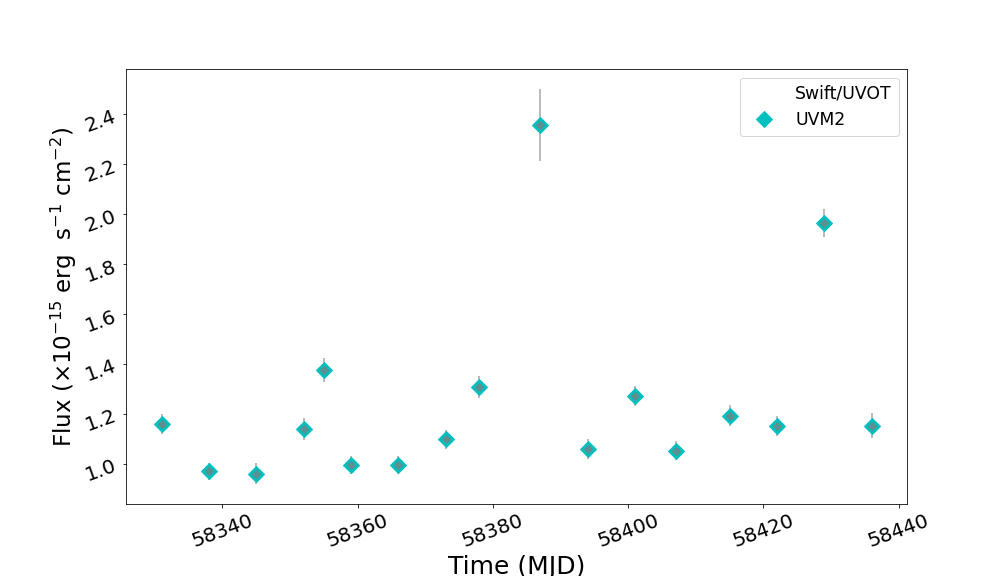}
\caption{{\it Swift}/UVOT lightcurve of the observations during weekly monitoring between August-November 2018.}
    \label{fig:UVOT}
\end{figure}

\begin{figure}
	\includegraphics[width=10.0cm]{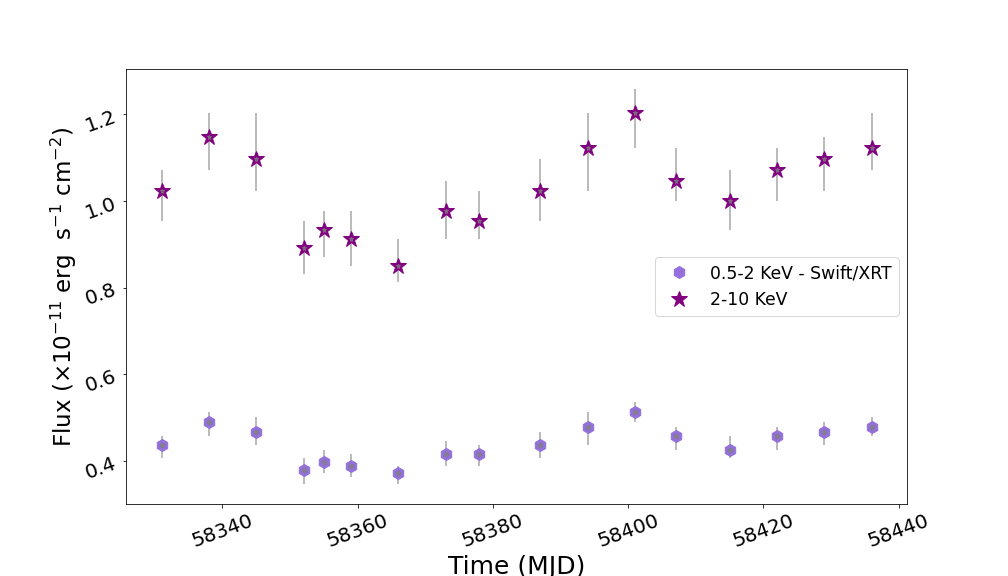}
\caption{\emph{Swift}/XRT lightcurves of PBC\,J2333.9-2343 in the soft (0.5-2 keV) and hard (2-10 keV) energy bands during the weekly monitoring between August-November 2018. }
    \label{fig:xraylightcurve}
\end{figure}

In this section we present the plots and tables of the monitoring campaigns with the different instruments. 
In Fig. \ref{fig:Eff} and Table \ref{Eff_fluxes}, results of the radio monitoring performed with the Effelsberg-100m single dish at 4.8, 8.5, 10.5, and 20.4 GHz are shown. Details about receivers and instrumental setup are reported in Table \ref{Effelsberg}. In Table \ref{VLBA} we report information about the VLBA observations.
Fig. \ref{fig:smarts} and Table \ref{tab:smarts} show fluxes of the NIR ($K$-band) and optical ($V$/$I$ bands) from the SMARTS-1.3m. Optical data are shown also in Fig. \ref{fig:ZTF} and Table \ref{tab:ztf} for ZTF in the gri filters, and in Fig. \ref{fig:ATLAS} and Table \ref{tab:atlas} for ATLAS in the oc filters. 
\emph{Swift} data is presented in Table \ref{tab:swift} and Figs. \ref{fig:UVOT} (UVM2 filter) and \ref{fig:xraylightcurve} (X-rays).

\vspace{1cm}


\begin{table*}
\centering
\caption{Flux densities between 4.8 and 20.4 GHz, collected at the different epochs (in MJD) with the Effelsberg-100m telescope. 
\label{Eff_fluxes}}
\begin{tabular}{ccccc}
\hline
	MJD			&	$S_{4.8}$			&	$S_{8.5}$ 			&	$S_{10.5}$			&	$S_{20.4}$ 		\\
				&	(Jy)				&	(Jy)				&	(Jy)				&	(Jy)			\\
\hline
58293	&	1.1$\pm$0.1		&	1.4$\pm$0.1		&	1.5$\pm$0.4		&	--				\\
58351	&	1.2$\pm$0.1		&	1.4$\pm$0.1		&	1.5$\pm$0.2		&	0.45$\pm$0.05 \\
58367	&	1.2$\pm$0.1		&	1.3$\pm$0.1		&	1.3$\pm$0.1		&	--		\\	
58398.9	&	1.2$\pm$0.1		&	1.3$\pm$0.1		&	1.5$\pm$0.2		&	0.96$\pm$0.09 \\
58411.9	&	1.2$\pm$0.1		&	1.4$\pm$0.1		&	1.5$\pm$0.2		&	2.0$\pm$0.2 \\
58451.7	&	1.1$\pm$0.1		&	1.2$\pm$0.1		&	1.3$\pm$0.1		&	1.2$\pm$0.1 \\
58478.7	&	1.1$\pm$0.1		&	1.4$\pm$0.1		&	1.6$\pm$0.2		&	1.8$\pm$0.2 \\
\hline
\end{tabular}
\end{table*}


\begin{table*}
\centering
\caption{Log of used receivers at the Effelsberg-100m telescope. \label{Effelsberg}}
\begin{tabular}{ccccc}
\hline
 	                               & S60mm 	    & S36mm 	& S28mm          &    S14mm      \\
\hline
Center frequency			     & 4.85 GHz	    & 8.35 GHz	& 10.45 GHz	     &    20.40 GHz  \\
Typical $T_{sys}$ (zenith)	         & 25 K		    & 22 K		& 50 K		     &    51 K       \\
HPBW				            & 145 arcsec	& 82 arcsec	& 66 arcsec	     &    40 arcsec  \\
Sensitivity				         & 1.55 K/Jy	    & 1.35 K/Jy	& 1.35 K/Jy	     &    1.07 K/Jy  \\
N. of pixels			                & 2			    & 1			& 2			     &    1          \\
\hline
\end{tabular}
\end{table*}


\begin{table*}
\centering
\caption{VLBA images parameters. Column 1: frequency, Col. 2: full width at half maximum (FWHM) with position angle, Col. 3: image noise, Col. 4: image peak brightness, Col. 6: total flux density.
\label{VLBA}}
\begin{tabular}{ccccc}
\hline
	$\nu$			&	FWHM			            &	RMS 			&	$I_\nu$			&	$S_\nu$ 	\\
	(GHz)			&	(mas$\times$mas, deg)	&	(mJy/beam)		&	(mJy/beam)		&	(Jy)		\\
\hline
    15              &  0.9$\times$0.3, --2.2         &   0.3            &  1072            &    1.7$\pm$0.2           \\  
    24              &  0.6$\times$0.2, --3.4         &   1.3            &   911            &    1.4$\pm$0.1           \\
\hline
\end{tabular}
\end{table*}


\onecolumn

\begin{center}
\begin{longtable}
{ccc}
\caption{A Log of the SMARTS-1.3m observations. This includes dates in MJD, and the fluxes in the optical and NIR bands (in mJy). In the optical, dates before MJD=58499 are in the $V$-filter, and after that in the $I$-filter.} \label{tab:smarts} \\

\hline \multicolumn{1}{c}{MJD} & \multicolumn{1}{c}{Flux ($V$/$I$) mJy} & \multicolumn{1}{c}{Flux ($K$) mJy} \\ \hline 
\endfirsthead

\multicolumn{3}{c}%
{{\bfseries \tablename\ \thetable{} -- continued from previous page}} \\
\hline \multicolumn{1}{c}{MJD} & \multicolumn{1}{c}{Flux ($V$/$I$) mJy} & \multicolumn{1}{c}{Flux ($K$) mJy} \\ \hline  
\endhead

\hline \multicolumn{3}{|r|}{{Continued on next page}} \\ \hline
\endfoot

\hline
\endlastfoot

58332.3	&	1.31	$\pm	$	0.02	&	8.69	$\pm	$	0.09	\\
58338.3	&	1.150	$\pm	$	0.005	&	9.0	$\pm	$	0.6	\\
58342.2	&	1.184	$\pm	$	0.003	&	8.9	$\pm	$	0.1	\\
58346.2	&	1.266	$\pm	$	0.008	&	10.02	$\pm	$	0.08	\\
58350.2	&	1.31	$\pm	$	0.02	&	9.7	$\pm	$	0.2	\\
58355.2	&	1.27	$\pm	$	0.02	&	8.7	$\pm	$	0.1	\\
58406.2	&	1.301	$\pm	$	0.005	&	9.6	$\pm	$	0.2	\\
58410.1	&	1.24	$\pm	$	0.03	&	9.32	$\pm	$	0.09	\\
58414.1	&	1.3 	$\pm	$	0.1	    &	8.7 	$\pm	$	0.3	\\
58418.1	&	1.35	$\pm	$	0.04	&	10.7	$\pm	$	0.9	\\
58422.2	&	1.266	$\pm	$	0.003	&	9.5	$\pm	$	0.3	\\
58426.1	&	1.180	$\pm	$	0.003	&	8.530	$\pm	$	0.007	\\
58430.1	&	1.50	$\pm	$	0.01	&	10.5	$\pm	$	0.3	\\
58434.1	&	1.260	$\pm	$	0.007	&	9.2	$\pm	$	0.2	\\
58438.1	&	1.16	$\pm	$	0.01	&	9.3	$\pm	$	0.4	\\
58442.1	&	1.136	$\pm	$	0.005	&	8.2	$\pm	$	0.2	\\
58447.1	&	1.24	$\pm	$	0.02	&	8.9	$\pm	$	0.3	\\
58451.1	&	1.212	$\pm	$	0.004	&	9.2	$\pm	$	0.2	\\
58455.1	&	1.128	$\pm	$	0.004	&	8.1	$\pm	$	0.2	\\
58459.1	&	1.157	$\pm	$	0.002	&	8.7	$\pm	$	0.3	\\
58463.1	&	1.093	$\pm	$	0.008	&	7.9	$\pm	$	0.4	\\
58467.0	&	1.140	$\pm	$	0.004	&	8.2	$\pm	$	0.2	\\
58471.1	&	1.050	$\pm	$	0.004	&	9.1	$\pm	$	0.3	\\
58475.1	&	1.014	$\pm	$	0.008	&	7.4	$\pm	$	0.3	\\
58479.1	&	1.048	$\pm	$	0.007	&	8.5	$\pm	$	0.5	\\
58483.0	&	1.040	$\pm	$	0.004	&	7.7	$\pm	$	0.2	\\
58487.0	&	1.083	$\pm	$	0.004	&	7.69	$\pm	$	0.05	\\
58491.0	&	1.120	$\pm	$	0.007	&	8.7	$\pm	$	0.2	\\
58495.0	&	1.12	$\pm	$	0.03	&	8.1	$\pm	$	0.2	\\
58499.0	&	1.12	$\pm	$	0.03	&	7.7	$\pm	$	0.4	\\
58606.4	&	2.29	$\pm	$	0.02	&	7.9	$\pm	$	0.2	\\
58607.4	&	2.30	$\pm	$	0.03	&	8.2	$\pm	$	0.1	\\
58608.4	&	2.42	$\pm	$	0.02	&	7.3	$\pm	$	0.3	\\
58609.4	&	2.44	$\pm	$	0.03	&	7.9	$\pm	$	0.2	\\
58611.4	&	2.77	$\pm	$	0.04	&	8.0	$\pm	$	0.2	\\
58612.4	&	2.763	$\pm	$	0.008	&	8.9	$\pm	$	0.4	\\
58613.4	&	2.4	    $\pm	$	0.1	&	8.8	$\pm	$	0.2	\\
58614.4	&	2.32	$\pm	$	0.03	&	8.1	$\pm	$	0.1	\\
58615.4	&	2.472	$\pm	$	0.007	&	8.48	$\pm	$	0.09	\\
58616.4	&	2.33	$\pm	$	0.01	&	8.7	$\pm	$	0.2	\\
58617.4	&	2.32	$\pm	$	0.01	&	8.2	$\pm	$	0.2	\\
58618.4	&	2.38	$\pm	$	0.03	&	8.2	$\pm	$	0.3	\\
58619.4	&	2.47	$\pm	$	0.04	&	8.1	$\pm	$	0.3	\\
58620.4	&	2.42	$\pm	$	0.02	&	8.4	$\pm	$	0.3	\\
58621.4	&	2.43	$\pm	$	0.01	&	7.6	$\pm	$	0.3	\\
58625.4	&	2.321	$\pm	$	0.008	&	7.69	$\pm	$	0.08	\\
58626.4	&	2.39	$\pm	$	0.02	&	8.8	$\pm	$	0.3	\\
58627.4	&	2.48	$\pm	$	0.07	&	7.48	$\pm	$	0.07	\\
58636.4	&	2.60	$\pm	$	0.01	&	7.5	$\pm	$	0.2	\\
58637.4	&	2.87	$\pm	$	0.02	&	8.0	$\pm	$	0.4	\\
58638.4	&	2.91	$\pm	$	0.02	&	9.1	$\pm	$	0.1	\\
58639.4	&	3.01	$\pm	$	0.03	&	9.3	$\pm	$	0.3	\\
58640.4	&	2.7 	$\pm	$	0.2	&	9.8	$\pm	$	0.3	\\
58641.3	&	3.16	$\pm	$	0.08	&	10.2	$\pm	$	0.6	\\
58642.4	&	2.85	$\pm	$	0.02	&	9.20	$\pm	$	0.02	\\
58643.4	&	3.35	$\pm	$	0.04	&	10.2	$\pm	$	0.2	\\
58644.4	&	3.48	$\pm	$	0.02	&	9.4	$\pm	$	0.1	\\
58645.4	&	3.64	$\pm	$	0.08	&	10.7	$\pm	$	0.1	\\
58646.4	&	2.90	$\pm	$	0.03	&	12.1	$\pm	$	0.3	\\
58647.4	&	3.01	$\pm	$	0.02	&	11.6	$\pm	$	0.3	\\
58655.3	&	2.31	$\pm	$	0.08	&	10.0	$\pm	$	0.1	\\
58656.4	&	2.31	$\pm	$	0.02	&	9.8	$\pm	$	0.1	\\
58657.4	&	2.30	$\pm	$	0.03	&	7.7	$\pm	$	0.3	\\
58659.4	&	2.43	$\pm	$	0.01	&	7.9	$\pm	$	0.3	\\
58660.4	&	2.43	$\pm	$	0.01	&	7.93	$\pm	$	0.07	\\
58662.4	&	2.56	$\pm	$	0.04	&	7.9	$\pm	$	0.2	\\
58664.3	&	2.58	$\pm	$	0.03	&	8.24	$\pm	$	0.08	\\
58665.4	&	2.629	$\pm	$	0.008	&	8.57	$\pm	$	0.06	\\
58666.4	&	2.70	$\pm	$	0.02	&	8.7	$\pm	$	0.1	\\
58668.4	&	2.53	$\pm	$	0.03	&	9.4	$\pm	$	0.1	\\
58669.4	&	2.384	$\pm	$	0.004	&	8.8	$\pm	$	0.3	\\
58670.4	&	2.43	$\pm	$	0.01	&	8.9	$\pm	$	0.3	\\
58672.4	&	2.73	$\pm	$	0.02	&	8.29	$\pm	$	0.04	\\
58673.4	&	2.70	$\pm	$	0.04	&	8.3	$\pm	$	0.2	\\
58674.4	&	2.56	$\pm	$	0.02	&	9.5	$\pm	$	0.1	\\
58678.4	&	2.540	$\pm	$	0.004	&	9.2	$\pm	$	0.1	\\
58679.4	&	2.642	$\pm	$	0.007	&	9.9	$\pm	$	0.1	\\
58680.4	&	2.67	$\pm	$	0.02	&	10.5	$\pm	$	0.4	\\
58681.4	&	2.86	$\pm	$	0.01	&	10.84	$\pm	$	0.08	\\
58683.4	&	2.90	$\pm	$	0.03	&	9.9	$\pm	$	0.2	\\
58684.4	&	2.78	$\pm	$	0.01	&	10.5	$\pm	$	0.2	\\
58687.3	&	2.55	$\pm	$	0.08	&	8.8	$\pm	$	0.1	\\
58689.3	&	2.476	$\pm	$	0.006	&	8.5	$\pm	$	0.1	\\
58690.4	&	2.43	$\pm	$	0.03	&	8.8	$\pm	$	0.2	\\
58691.3	&	2.426	$\pm	$	0.006	&	8.8	$\pm	$	0.1	\\
58692.3	&	2.40	$\pm	$	0.01	&	8.68	$\pm	$	0.09	\\
58693.3	&	2.41	$\pm	$	0.01	&	8.5	$\pm	$	0.2	\\
58694.3	&	2.312	$\pm	$	0.005	&	8.4	$\pm	$	0.1	\\
58695.3	&	2.27	$\pm	$	0.02	&	8.1	$\pm	$	0.4	\\
\end{longtable}
\end{center}

\begin{center}
\begin{longtable}
{ccc}
\caption{ Log of the ZTF  forced photometry observations. Include the dates in MJD, filter, and color corrected apparent magnitudes.} \label{tab:ztf} \\

\hline \multicolumn{1}{c}{MJD} & \multicolumn{1}{c}{Filter} & \multicolumn{1}{c}{Magnitude (mag)} \\ \hline 
\endfirsthead

\multicolumn{3}{c}%
{{\bfseries \tablename\ \thetable{} -- continued from previous page}} \\
\hline \multicolumn{1}{c}{MJD} & \multicolumn{1}{c}{Filter} & \multicolumn{1}{c}{Magnitude (mag)} \\ \hline 
\endhead

\hline \multicolumn{3}{|r|}{{Continued on next page}} \\ \hline
\endfoot

\hline
\endlastfoot

58322.439	&	g	&	16.969$\pm	$0.006	\\
58322.481	&	r	&	16.215$\pm	$0.004	\\
58325.417	&	g	&	17.01$\pm	$0.03	\\
58333.399	&	i	&	15.831$\pm	$0.005	\\
58334.418	&	r	&	16.041$\pm	$0.005	\\
58334.46	&	g	&	16.71$\pm	$0.01	\\
58337.396	&	g	&	16.867$\pm	$0.006	\\
58337.460	&	i	&	15.869$\pm	$0.005	\\
58338.408	&	g	&	16.908$\pm	$0.006	\\
58338.463	&	r	&	16.217$\pm	$0.006	\\
58342.416	&	g	&	16.878$\pm	$0.006	\\
58342.501	&	r	&	16.186$\pm	$0.006	\\
58342.502	&	r	&	16.194$\pm	$0.008	\\
58344.397	&	g	&	16.885$\pm	$0.005	\\
58344.420	&	r	&	16.174$\pm	$0.004	\\
58345.376	&	r	&	16.163$\pm	$0.004	\\
58347.376	&	i	&	15.769$\pm	$0.005	\\
58347.398	&	r	&	16.079$\pm	$0.003	\\
58348.369	&	g	&	16.757$\pm	$0.005	\\
58348.370	&	g	&	16.760$\pm	$0.005	\\
58348.398	&	r	&	16.079$\pm	$0.003	\\
58348.398	&	r	&	16.092$\pm	$0.003	\\
58350.377	&	r	&	16.101$\pm	$0.003	\\
58351.399	&	g	&	16.770$\pm	$0.004	\\
58351.400	&	g	&	16.781$\pm	$0.005	\\
58352.355	&	g	&	16.75$\pm	$0.01	\\
58354.377	&	r	&	16.148$\pm	$0.008	\\
58355.349	&	r	&	16.11$\pm	$0.01	\\
58356.382	&	r	&	16.05$\pm	$0.01	\\
58359.367	&	r	&	16.09$\pm	$0.01	\\
58359.368	&	r	&	16.19$\pm	$0.01	\\
58360.372	&	g	&	16.92$\pm	$0.02	\\
58361.354	&	r	&	16.113$\pm	$0.007	\\
58361.356	&	i	&	15.814$\pm	$0.006	\\
58362.334	&	r	&	16.036$\pm	$0.004	\\
58362.335	&	r	&	16.035$\pm	$0.004	\\
58363.337	&	g	&	16.799$\pm	$0.009	\\
58363.423	&	r	&	16.116$\pm	$0.005	\\
58364.349	&	r	&	16.080$\pm	$0.005	\\
58364.398	&	g	&	16.708$\pm	$0.009	\\
58365.326	&	i	&	15.806$\pm	$0.005	\\
58365.355	&	r	&	16.126$\pm	$0.005	\\
58365.356	&	r	&	16.125$\pm	$0.004	\\
58365.398	&	g	&	16.807$\pm	$0.007	\\
58366.350	&	r	&	16.061$\pm	$0.003	\\
58367.336	&	r	&	16.110$\pm	$0.004	\\
58367.397	&	g	&	16.799$\pm	$0.005	\\
58368.334	&	r	&	16.161$\pm	$0.004	\\
58368.335	&	r	&	16.168$\pm	$0.003	\\
58368.356	&	g	&	16.825$\pm	$0.005	\\
58369.331	&	g	&	16.811$\pm	$0.006	\\
58369.357	&	r	&	16.121$\pm	$0.005	\\
58370.307	&	r	&	16.110$\pm	$0.005	\\
58370.350	&	g	&	16.779$\pm	$0.006	\\
58371.320	&	r	&	16.167$\pm	$0.003	\\
58371.322	&	r	&	16.185$\pm	$0.003	\\
58371.335	&	g	&	16.829$\pm	$0.006	\\
58371.352	&	g	&	16.856$\pm	$0.005	\\
58372.373	&	r	&	16.129$\pm	$0.004	\\
58372.400	&	g	&	16.809$\pm	$0.007	\\
58373.309	&	r	&	16.145$\pm	$0.005	\\
58373.335	&	i	&	15.835$\pm	$0.007	\\
58373.358	&	g	&	16.841$\pm	$0.007	\\
58374.334	&	r	&	16.145$\pm	$0.004	\\
58374.335	&	r	&	16.148$\pm	$0.004	\\
58374.414	&	g	&	16.829$\pm	$0.008	\\
58374.415	&	g	&	16.840$\pm	$0.008	\\
58375.354	&	r	&	16.159$\pm	$0.004	\\
58376.312	&	g	&	16.839$\pm	$0.006	\\
58376.337	&	r	&	16.141$\pm	$0.004	\\
58377.293	&	r	&	16.163$\pm	$0.005	\\
58377.310	&	r	&	16.170$\pm	$0.004	\\
58377.337	&	g	&	16.884$\pm	$0.007	\\
58377.338	&	g	&	16.870$\pm	$0.007	\\
58378.287	&	r	&	16.182$\pm	$0.005	\\
58378.314	&	g	&	16.881$\pm	$0.007	\\
58379.313	&	r	&	16.141$\pm	$0.005	\\
58380.293	&	r	&	16.093$\pm	$0.005	\\
58380.294	&	r	&	16.070$\pm	$0.005	\\
58380.337	&	g	&	16.726$\pm	$0.008	\\
58382.273	&	i	&	15.820$\pm	$0.006	\\
58382.314	&	r	&	16.101$\pm	$0.008	\\
58383.272	&	g	&	16.80$\pm	$0.02	\\
58383.274	&	g	&	16.83$\pm	$0.02	\\
58383.294	&	r	&	16.12$\pm	$0.01	\\
58383.295	&	r	&	16.08$\pm	$0.01	\\
58386.293	&	r	&	15.705$\pm	$0.009	\\
58386.298	&	r	&	15.68$\pm	$0.01	\\
58387.355	&	i	&	15.567$\pm	$0.009	\\
58388.271	&	r	&	15.969$\pm	$0.008	\\
58389.265	&	r	&	15.922$\pm	$0.006	\\
58389.266	&	r	&	15.940$\pm	$0.006	\\
58389.343	&	g	&	16.631$\pm	$0.013	\\
58390.270	&	r	&	16.074$\pm	$0.006	\\
58390.300	&	g	&	16.73$\pm	$0.01	\\
58397.283	&	r	&	16.079$\pm	$0.004	\\
58397.284	&	r	&	16.077$\pm	$0.004	\\
58400.231	&	i	&	15.785$\pm	$0.005	\\
58423.214	&	g	&	16.81$\pm	$0.01	\\
58423.214	&	g	&	16.79$\pm	$0.01	\\
58425.186	&	g	&	16.841$\pm	$0.009	\\
58426.226	&	g	&	16.82$\pm	$0.01	\\
58426.234	&	r	&	16.143$\pm	$0.009	\\
58426.235	&	r	&	16.157$\pm	$0.009	\\
58428.187	&	g	&	16.597$\pm	$0.006	\\
58428.250	&	r	&	15.933$\pm	$0.005	\\
58429.23	&	g	&	16.436$\pm	$0.006	\\
58429.247	&	g	&	16.453$\pm	$0.006	\\
58429.254	&	r	&	15.841$\pm	$0.004	\\
58429.255	&	r	&	15.841$\pm	$0.004	\\
58430.210	&	i	&	15.704$\pm	$0.006	\\
58430.262	&	g	&	16.582$\pm	$0.007	\\
58431.188	&	r	&	15.809$\pm	$0.008	\\
58432.231	&	g	&	16.610$\pm	$0.009	\\
58432.243	&	g	&	16.638$\pm	$0.009	\\
58432.254	&	r	&	15.945$\pm	$0.007	\\
58432.255	&	r	&	15.986$\pm	$0.007	\\
58434.210	&	i	&	15.821$\pm	$0.009	\\
58457.171	&	r	&	16.20$\pm	$0.01	\\
58472.165	&	g	&	17.10$\pm	$0.03	\\

\end{longtable}
\end{center}

\begin{center}
\begin{longtable}
{lll}
\caption{Log of the ATLAS forced photometry observations. Includes dates in MJD, filter, and magnitudes.} \label{tab:atlas} \\

\hline \multicolumn{1}{c}{MJD} & \multicolumn{1}{c}{Filter} & \multicolumn{1}{c}{Magnitude (mag)} \\ \hline 
\endfirsthead

\multicolumn{3}{c}%
{{\bfseries \tablename\ \thetable{} -- continued from previous page}} \\
\hline \multicolumn{1}{c}{MJD} & \multicolumn{1}{c}{Filter} & \multicolumn{1}{c}{Magnitude (mag)} \\ \hline 
\endhead

\hline \multicolumn{3}{|r|}{{Continued on next page}} \\ \hline
\endfoot

\hline
\endlastfoot
58321.519	&	o	&	15.76$\pm	$0.01	\\
58321.527	&	o	&	15.78$\pm	$0.01	\\
58321.535	&	o	&	15.78$\pm	$0.01	\\
58321.542	&	o	&	15.75$\pm	$0.01	\\
58325.452	&	o	&	15.86$\pm	$0.02	\\
58325.460	&	o	&	15.90$\pm	$0.02	\\
58325.469	&	o	&	15.89$\pm	$0.02	\\
58325.477	&	o	&	15.84$\pm	$0.02	\\
58331.508	&	o	&	15.65$\pm	$0.03	\\
58331.511	&	o	&	15.68$\pm	$0.03	\\
58331.524	&	o	&	15.62$\pm	$0.02	\\
58331.533	&	o	&	15.60$\pm	$0.02	\\
58333.608	&	c	&	16.22$\pm	$0.02	\\
58333.616	&	c	&	16.23$\pm	$0.02	\\
58333.625	&	c	&	16.25$\pm	$0.02	\\
58337.497	&	c	&	16.23$\pm	$0.01	\\
58337.510	&	c	&	16.19$\pm	$0.01	\\
58337.512	&	c	&	16.25$\pm	$0.01	\\
58337.521	&	c	&	16.26$\pm	$0.01	\\
58341.489	&	c	&	16.20$\pm	$0.01	\\
58341.498	&	c	&	16.20$\pm	$0.01	\\
58341.507	&	c	&	16.22$\pm	$0.01	\\
58341.509	&	c	&	16.20$\pm	$0.01	\\
58341.512	&	c	&	16.18$\pm	$0.01	\\
58341.515	&	c	&	16.23$\pm	$0.01	\\
58341.521	&	c	&	16.30$\pm	$0.01	\\
58343.486	&	o	&	15.46$\pm	$0.01	\\
58343.497	&	o	&	15.44$\pm	$0.01	\\
58343.499	&	o	&	15.48$\pm	$0.01	\\
58343.501	&	o	&	15.47$\pm	$0.01	\\
58343.503	&	o	&	15.35$\pm	$0.01	\\
58343.510	&	o	&	15.45$\pm	$0.01	\\
58343.511	&	o	&	15.34$\pm	$0.01	\\
58343.520	&	o	&	15.49$\pm	$0.01	\\
58345.486	&	c	&	16.35$\pm	$0.04	\\
58345.492	&	c	&	16.34$\pm	$0.04	\\
58345.499	&	c	&	16.31$\pm	$0.03	\\
58345.516	&	c	&	16.20$\pm	$0.05	\\
58345.517	&	c	&	16.31$\pm	$0.05	\\
58347.467	&	o	&	15.48$\pm	$0.01	\\
58347.476	&	o	&	15.52$\pm	$0.01	\\
58347.491	&	o	&	15.53$\pm	$0.01	\\
58347.500	&	o	&	15.540$\pm	$0.009	\\
58349.566	&	o	&	15.63$\pm	$0.01	\\
58351.518	&	o	&	15.40$\pm	$0.01	\\
58351.520	&	o	&	15.46$\pm	$0.01	\\
58351.528	&	o	&	15.50$\pm	$0.01	\\
58351.530	&	o	&	15.44$\pm	$0.01	\\
58351.538	&	o	&	15.50$\pm	$0.01	\\
58351.548	&	o	&	15.50$\pm	$0.01	\\
58361.452	&	o	&	15.71$\pm	$0.01	\\
58361.455	&	o	&	15.70$\pm	$0.01	\\
58361.464	&	o	&	15.74$\pm	$0.01	\\
58361.465	&	o	&	15.71$\pm	$0.01	\\
58363.437	&	o	&	15.55$\pm	$0.01	\\
58363.453	&	o	&	15.61$\pm	$0.01	\\
58363.463	&	o	&	15.45$\pm	$0.01	\\
58363.466	&	o	&	15.56$\pm	$0.01	\\
58367.436	&	o	&	15.551$\pm	$0.009	\\
58367.453	&	o	&	15.551$\pm	$0.009	\\
58367.461	&	o	&	15.538$\pm	$0.009	\\
58367.463	&	o	&	15.568$\pm	$0.009	\\
58369.431	&	o	&	15.77$\pm	$0.01	\\
58369.439	&	o	&	15.74$\pm	$0.01	\\
58369.440	&	o	&	15.74$\pm	$0.01	\\
58369.454	&	o	&	15.74$\pm	$0.01	\\
58371.429	&	o	&	15.49$\pm	$0.01	\\
58371.430	&	o	&	15.54$\pm	$0.01	\\
58371.441	&	o	&	15.61$\pm	$0.01	\\
58371.456	&	o	&	15.54$\pm	$0.01	\\
58375.426	&	o	&	15.546$\pm	$0.009	\\
58375.435	&	o	&	15.60$\pm	$0.01	\\
58375.436	&	o	&	15.52$\pm	$0.01	\\
58377.467	&	o	&	15.74$\pm	$0.01	\\
58377.478	&	o	&	15.75$\pm	$0.01	\\
58377.487	&	o	&	15.73$\pm	$0.01	\\
58377.498	&	o	&	15.76$\pm	$0.01	\\
58379.458	&	o	&	15.50$\pm	$0.02	\\
58379.468	&	o	&	15.57$\pm	$0.02	\\
58379.478	&	o	&	15.59$\pm	$0.02	\\
58379.488	&	o	&	15.58$\pm	$0.02	\\
58381.437	&	o	&	15.65$\pm	$0.01	\\
58381.447	&	o	&	15.63$\pm	$0.01	\\
58381.456	&	o	&	15.59$\pm	$0.01	\\
58381.466	&	o	&	15.61$\pm	$0.01	\\
58391.389	&	o	&	15.49$\pm	$0.01	\\
58391.390	&	o	&	15.52$\pm	$0.01	\\
58391.404	&	o	&	15.49$\pm	$0.01	\\
58391.416	&	o	&	15.52$\pm	$0.01	\\
58393.385	&	o	&	15.66$\pm	$0.01	\\
58393.385	&	o	&	15.69$\pm	$0.01	\\
58393.399	&	o	&	15.68$\pm	$0.01	\\
58393.411	&	o	&	15.59$\pm	$0.01	\\
58395.384	&	o	&	15.56$\pm	$0.01	\\
58395.389	&	o	&	15.60$\pm	$0.01	\\
58395.403	&	o	&	15.56$\pm	$0.01	\\
58401.370	&	c	&	16.11$\pm	$0.01	\\
58401.379	&	c	&	16.07$\pm	$0.01	\\
58401.387	&	c	&	16.12$\pm	$0.01	\\
58401.393	&	c	&	16.07$\pm	$0.01	\\
58403.414	&	o	&	15.50$\pm	$0.02	\\
58403.424	&	o	&	15.51$\pm	$0.01	\\
58403.434	&	o	&	15.50$\pm	$0.01	\\
58403.444	&	o	&	15.50$\pm	$0.01	\\
58405.406	&	c	&	16.18$\pm	$0.01	\\
58405.416	&	c	&	16.17$\pm	$0.01	\\
58405.425	&	c	&	16.16$\pm	$0.01	\\
58405.435	&	c	&	16.13$\pm	$0.01	\\
58417.376	&	o	&	15.64$\pm	$0.01	\\
58417.385	&	o	&	15.65$\pm	$0.01	\\
58417.395	&	o	&	15.62$\pm	$0.01	\\
58417.404	&	o	&	15.61$\pm	$0.01	\\
58423.398	&	o	&	15.46$\pm	$0.01	\\
58423.408	&	o	&	15.44$\pm	$0.01	\\
58427.358	&	o	&	15.41$\pm	$0.01	\\
58427.367	&	o	&	15.39$\pm	$0.01	\\
58427.378	&	o	&	15.38$\pm	$0.01	\\
58427.387	&	o	&	15.38$\pm	$0.01	\\
58429.358	&	c	&	16.02$\pm	$0.01	\\
58429.368	&	c	&	16.04$\pm	$0.01	\\
58429.377	&	c	&	15.99$\pm	$0.01	\\
58429.386	&	c	&	16.03$\pm	$0.01	\\
58431.346	&	o	&	15.247$\pm	$0.009	\\
58431.356	&	o	&	15.232$\pm	$0.008	\\
58431.366	&	o	&	15.222$\pm	$0.008	\\
58431.375	&	o	&	15.201$\pm	$0.008	\\
58433.349	&	c	&	16.17$\pm	$0.02	\\
58433.358	&	c	&	16.13$\pm	$0.02	\\
58433.377	&	c	&	16.18$\pm	$0.02	\\
58435.316	&	o	&	15.47$\pm	$0.01	\\
58435.326	&	o	&	15.47$\pm	$0.01	\\
58435.331	&	o	&	15.51$\pm	$0.01	\\
58435.335	&	o	&	15.46$\pm	$0.01	\\
58435.340	&	o	&	15.49$\pm	$0.01	\\
58435.345	&	o	&	15.48$\pm	$0.01	\\
58435.350	&	o	&	15.51$\pm	$0.01	\\
58435.359	&	o	&	15.49$\pm	$0.01	\\
58443.317	&	o	&	15.55$\pm	$0.02	\\
58443.326	&	o	&	15.61$\pm	$0.02	\\
58443.335	&	o	&	15.59$\pm	$0.02	\\
58443.344	&	o	&	15.65$\pm	$0.02	\\
58447.326	&	o	&	15.56$\pm	$0.01	\\
58447.335	&	o	&	15.56$\pm	$0.02	\\
58447.344	&	o	&	15.54$\pm	$0.02	\\
58447.355	&	o	&	15.54$\pm	$0.02	\\
58449.305	&	o	&	15.73$\pm	$0.01	\\
58449.314	&	o	&	15.71$\pm	$0.01	\\
58449.324	&	o	&	15.74$\pm	$0.01	\\
58449.333	&	o	&	15.75$\pm	$0.01	\\
58451.300	&	o	&	15.54$\pm	$0.01	\\
58451.309	&	o	&	15.264$\pm	$0.008	\\
58451.319	&	o	&	15.51$\pm	$0.01	\\
58451.328	&	o	&	15.460$\pm	$0.009	\\
58453.310	&	o	&	15.85$\pm	$0.02	\\
58453.319	&	o	&	15.80$\pm	$0.01	\\
58453.329	&	o	&	15.83$\pm	$0.01	\\
58453.338	&	o	&	15.82$\pm	$0.01	\\
58455.286	&	o	&	15.67$\pm	$0.01	\\
58455.294	&	o	&	15.67$\pm	$0.01	\\
58455.302	&	o	&	15.65$\pm	$0.01	\\
58455.310	&	o	&	15.70$\pm	$0.01	\\
58459.260	&	o	&	15.59$\pm	$0.01	\\
58459.268	&	o	&	15.559$\pm	$0.009	\\
58459.276	&	o	&	15.59$\pm	$0.01	\\
58459.284	&	o	&	15.61$\pm	$0.01	\\
58463.248	&	o	&	15.57$\pm	$0.01	\\
58463.256	&	o	&	15.54$\pm	$0.01	\\
58463.264	&	o	&	15.56$\pm	$0.01	\\
58463.272	&	o	&	15.47$\pm	$0.01	\\
58465.236	&	o	&	15.77$\pm	$0.01	\\
58465.243	&	o	&	15.79$\pm	$0.01	\\
58465.250	&	o	&	15.80$\pm	$0.01	\\
58465.258	&	o	&	15.83$\pm	$0.01	\\
58467.249	&	o	&	15.70$\pm	$0.02	\\
58467.256	&	o	&	15.74$\pm	$0.02	\\
58467.263	&	o	&	15.68$\pm	$0.02	\\
58467.270	&	o	&	15.65$\pm	$0.02	\\
58469.251	&	o	&	15.76$\pm	$0.01	\\
58469.259	&	o	&	15.76$\pm	$0.02	\\
58469.267	&	o	&	15.76$\pm	$0.01	\\
58469.274	&	o	&	15.75$\pm	$0.02	\\
58471.243	&	o	&	15.68$\pm	$0.02	\\
58471.250	&	o	&	15.67$\pm	$0.02	\\
58471.257	&	o	&	15.63$\pm	$0.02	\\
58471.264	&	o	&	15.63$\pm	$0.02	\\
58473.243	&	o	&	15.78$\pm	$0.02	\\
58473.250	&	o	&	15.78$\pm	$0.02	\\
58473.257	&	o	&	15.68$\pm	$0.02	\\
58473.264	&	o	&	15.65$\pm	$0.02	\\
58474.303	&	o	&	15.86$\pm	$0.03	\\
58474.311	&	o	&	15.85$\pm	$0.03	\\
58474.319	&	o	&	15.86$\pm	$0.02	\\
58475.217	&	o	&	15.57$\pm	$0.02	\\
58475.226	&	o	&	15.66$\pm	$0.02	\\
58475.235	&	o	&	15.62$\pm	$0.02	\\
58475.244	&	o	&	15.66$\pm	$0.03	\\
58479.225	&	o	&	15.61$\pm	$0.01	\\
58479.234	&	o	&	15.62$\pm	$0.01	\\
58479.241	&	o	&	15.56$\pm	$0.01	\\
58479.248	&	o	&	15.59$\pm	$0.01	\\
58485.231	&	c	&	16.31$\pm	$0.01	\\
58485.240	&	c	&	16.26$\pm	$0.02	\\
58485.248	&	c	&	16.26$\pm	$0.01	\\
58485.257	&	c	&	16.29$\pm	$0.02	\\
58487.221	&	o	&	15.60$\pm	$0.01	\\
58487.231	&	o	&	15.58$\pm	$0.01	\\
58487.239	&	o	&	15.60$\pm	$0.01	\\
58487.248	&	o	&	15.56$\pm	$0.01	\\
58489.224	&	c	&	16.15$\pm	$0.02	\\
58489.234	&	c	&	16.19$\pm	$0.02	\\
58489.243	&	c	&	16.14$\pm	$0.02	\\
58489.252	&	c	&	16.12$\pm	$0.02	\\
58491.207	&	o	&	15.61$\pm	$0.02	\\
58491.213	&	o	&	15.58$\pm	$0.01	\\
58491.221	&	o	&	15.58$\pm	$0.01	\\
58491.230	&	o	&	15.55$\pm	$0.01	\\
58493.212	&	c	&	16.28$\pm	$0.02	\\
58493.218	&	c	&	16.28$\pm	$0.02	\\
58493.226	&	c	&	16.30$\pm	$0.02	\\
58493.235	&	c	&	16.29$\pm	$0.02	\\
58495.213	&	o	&	15.67$\pm	$0.02	\\
58495.221	&	o	&	15.71$\pm	$0.01	\\
58495.230	&	o	&	15.63$\pm	$0.01	\\
58495.240	&	o	&	15.62$\pm	$0.01	\\
58497.210	&	o	&	15.73$\pm	$0.02	\\
58497.216	&	o	&	15.81$\pm	$0.02	\\
58497.224	&	o	&	15.80$\pm	$0.01	\\
58497.233	&	o	&	15.82$\pm	$0.01	\\
58499.211	&	o	&	15.63$\pm	$0.02	\\
58499.218	&	o	&	15.56$\pm	$0.02	\\
58501.220	&	o	&	15.67$\pm	$0.02	\\
58501.227	&	o	&	15.81$\pm	$0.02	\\
58507.214	&	o	&	15.63$\pm	$0.02	\\
58623.607	&	o	&	15.66$\pm	$0.03	\\
58623.613	&	o	&	15.69$\pm	$0.03	\\
58625.596	&	o	&	15.83$\pm	$0.03	\\
58625.600	&	o	&	15.83$\pm	$0.03	\\
58625.607	&	o	&	15.80$\pm	$0.02	\\
58645.597	&	o	&	15.340$\pm	$0.008	\\
58645.602	&	o	&	15.337$\pm	$0.008	\\
58645.612	&	o	&	15.310$\pm	$0.009	\\
58645.615	&	o	&	15.31$\pm	$0.01	\\
58647.580	&	o	&	15.36$\pm	$0.01	\\
58647.584	&	o	&	15.42$\pm	$0.02	\\
58647.590	&	o	&	15.31$\pm	$0.01	\\
58647.617	&	o	&	15.45$\pm	$0.02	\\
58649.531	&	o	&	15.53$\pm	$0.01	\\
58649.535	&	o	&	15.54$\pm	$0.02	\\
58649.540	&	o	&	15.52$\pm	$0.01	\\
58649.551	&	o	&	15.52$\pm	$0.01	\\
58651.561	&	o	&	15.51$\pm	$0.02	\\
58651.564	&	o	&	15.56$\pm	$0.02	\\
58651.569	&	o	&	15.53$\pm	$0.02	\\
58651.580	&	o	&	15.50$\pm	$0.02	\\
58659.601	&	o	&	15.67$\pm	$0.01	\\
58659.609	&	o	&	15.69$\pm	$0.01	\\
58659.615	&	o	&	15.67$\pm	$0.02	\\
58665.618	&	c	&	16.25$\pm	$0.02	\\
58665.621	&	c	&	16.24$\pm	$0.04	\\
58667.603	&	o	&	15.542$\pm	$0.009	\\
58667.608	&	o	&	15.571$\pm	$0.009	\\
58667.612	&	o	&	15.56$\pm	$0.01	\\
58667.621	&	o	&	15.63$\pm	$0.02	\\
58669.600	&	c	&	16.23$\pm	$0.01	\\
58669.603	&	c	&	16.29$\pm	$0.01	\\
58669.614	&	c	&	16.30$\pm	$0.01	\\
58669.620	&	c	&	16.31$\pm	$0.02	\\
58670.602	&	c	&	16.25$\pm	$0.01	\\
58670.612	&	c	&	16.28$\pm	$0.01	\\
58670.626	&	c	&	16.31$\pm	$0.04	\\
58671.603	&	c	&	16.26$\pm	$0.01	\\
58671.607	&	c	&	16.29$\pm	$0.01	\\
58671.613	&	c	&	16.22$\pm	$0.01	\\
58671.618	&	c	&	16.29$\pm	$0.02	\\
58674.606	&	o	&	15.76$\pm	$0.01	\\
58674.610	&	o	&	15.75$\pm	$0.01	\\
58674.615	&	o	&	15.76$\pm	$0.01	\\
58674.618	&	o	&	15.78$\pm	$0.01	\\
58679.455	&	o	&	15.62$\pm	$0.03	\\
58679.459	&	o	&	15.58$\pm	$0.03	\\
58679.465	&	o	&	15.49$\pm	$0.02	\\
58679.477	&	o	&	15.57$\pm	$0.03	\\
58689.596	&	o	&	15.77$\pm	$0.01	\\
58689.598	&	o	&	15.82$\pm	$0.01	\\
58689.603	&	o	&	15.80$\pm	$0.01	\\
58689.615	&	o	&	15.79$\pm	$0.01	\\
58691.571	&	o	&	15.56$\pm	$0.01	\\
58691.577	&	o	&	15.62$\pm	$0.01	\\
58691.580	&	o	&	15.56$\pm	$0.01	\\
58691.581	&	o	&	15.68$\pm	$0.01	\\
58691.595	&	o	&	15.56$\pm	$0.01	\\
58693.448	&	c	&	16.33$\pm	$0.01	\\
58693.448	&	c	&	16.34$\pm	$0.01	\\
58693.449	&	c	&	16.34$\pm	$0.01	\\
58693.460	&	c	&	16.37$\pm	$0.01	\\
58695.608	&	o	&	15.60$\pm	$0.01	\\
58695.613	&	o	&	15.62$\pm	$0.01	\\
58695.619	&	o	&	15.61$\pm	$0.01	\\
58695.621	&	o	&	15.69$\pm	$0.01	\\
58695.625	&	o	&	15.66$\pm	$0.01	\\
58699.591	&	o	&	15.65$\pm	$0.01	\\
58699.594	&	o	&	15.64$\pm	$0.01	\\
58699.601	&	o	&	15.60$\pm	$0.01	\\
58699.614	&	o	&	15.62$\pm	$0.01	\\
58699.620	&	o	&	15.71$\pm	$0.01	\\

\end{longtable}
\end{center}

\begin{table*}
\centering
\caption{Log of the \emph{Swift} observations. Includes the observation ID (ObsID), dates in MJD, exposure times (in ksec), X-ray fluxes in the 0.5-2 keV and 2-10 keV energy bands, and UV flux in the UVM2 filter.  \label{tab:swift}}
\begin{tabular}{cccccc}
\hline
 ObsID	&	MJD	&	Exposure Time	&	Flux (0.5-2 keV) 	&	Flux (2-10 keV)	&	Flux (UVM2)	\\ 
  &   & (ks) & (10$^{-11}$ erg s$^{-1}$cm$^{-2}$) & (10$^{-11}$ erg s$^{-1}$cm$^{-2}$) & (10$^{-15}$ erg s$^{-1}$cm$^{-2}$) \\
  \hline
00031731014	&	58331.0	&	2.7	&	0.44$_{-0.03}^{+0.02}	$ &	1.02$_{-0.07}^{+0.05}	$ &	1.16$\pm	$0.04\\
00031731015	&	58338.0	&	2.9	&	0.49$_{-0.03}^{+0.02}	$ &	1.15$_{-0.08}^{+0.05}	$ &	0.97$\pm	$0.03\\
00031731016	&	58345.0	&	1.7	&	0.47$_{-0.03}^{+0.03}	$ &	1.10$_{-0.07}^{+0.11}	$ &	0.96$\pm	$0.04\\
00031731017	&	58352.0	&	2.5	&	0.38$_{-0.03}^{+0.03}	$ &	0.89$_{-0.06}^{+0.06}	$ &	1.14$\pm	$0.04\\
00031731018	&	58355.0	&	2.5	&	0.40$_{-0.03}^{+0.03}	$ &	0.93$_{-0.06}^{+0.04}	$ &	1.38$\pm	$0.05\\
00031731019	&	58359.0	&	2.9	&	0.39$_{-0.03}^{+0.03}	$ &	0.91$_{-0.06}^{+0.07}	$ &	1.00$\pm	$0.04\\
00031731020	&	58366.0	&	3.1	&	0.37$_{-0.02}^{+0.02}	$ &	0.85$_{-0.04}^{+0.06}	$ &	1.00$\pm	$0.04\\
00031731021	&	58373.0	&	2.9	&	0.42$_{-0.03}^{+0.03}	$ &	0.98$_{-0.07}^{+0.07}	$ &	1.09$\pm	$0.04\\
00031731022	&	58378.0	&	2.8	&	0.42$_{-0.03}^{+0.02}	$ &	0.95$_{-0.04}^{+0.07}	$ &	1.31$\pm	$0.04\\
00031731023	&	58387.0	&	2.3	&	0.44$_{-0.03}^{+0.03}	$ &	1.02$_{-0.07}^{+0.07}	$ &	2.4$\pm	$0.1\\
00031731024	&	58394.0	&	1.2	&	0.48$_{-0.04}^{+0.03}	$ &	1.12$_{-0.10}^{+0.08}	$ &	1.06$\pm	$0.04\\
00031731025	&	58401.0	&	2.9	&	0.51$_{-0.02}^{+0.02}	$ &	1.20$_{-0.08}^{+0.06}	$ &	1.27$\pm	$0.04\\
00031731026	&	58407.0	&	2.7	&	0.46$_{-0.03}^{+0.02}	$ &	1.05$_{-0.05}^{+0.07}	$ &	1.05$\pm	$0.04\\
00031731027	&	58415.0	&	2.9	&	0.43$_{-0.02}^{+0.03}	$ &	1.00$_{-0.07}^{+0.07}	$ &	1.19$\pm	$0.04\\
00031731028	&	58422.0	&	3.1	&	0.46$_{-0.03}^{+0.02}	$ &	1.07$_{-0.07}^{+0.05}	$ &	1.15$\pm	$0.04\\
00031731029	&	58429.0	&	2.8	&	0.47$_{-0.03}^{+0.02}	$ &	1.10$_{-0.07}^{+0.05}	$ &	1.96$\pm	$0.06\\
00031731030	&	58436.0	&	2.8	&	0.48$_{-0.02}^{+0.02}	$ &	1.12$_{-0.05}^{+0.08}	$ &	1.15$\pm	$0.05\\
\hline
\end{tabular}
\end{table*}

\twocolumn

\section{Flare fitting}\label{appendixb}

\begin{figure}
	\includegraphics[width=10.0cm]{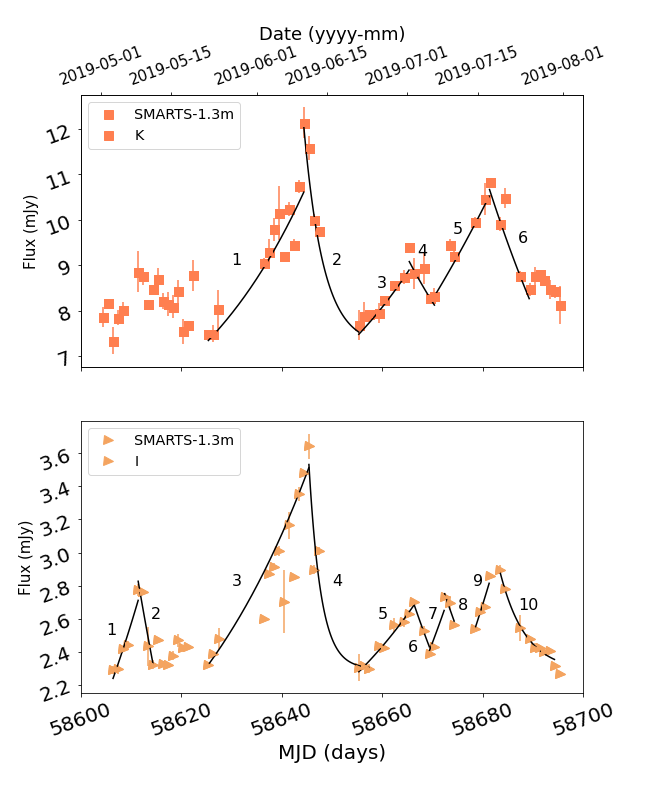}
\caption{Flux light curves in $I$ and $K$ bands from SMARTS-1.3m. Each of the segments are fitted by a power law to estimate the doubling/halving time that can be found in Table \ref{tab:flares}. }
    \label{fig:flares}
\end{figure}

We used the SMARTS-1.3m light curves from 2019, the ones with higher cadence, to estimate variability timescales. We fit different segments of the light curves with clear upwards or downwards trends, using power-laws of the form $flux = A(\lambda / B)^C + D$, where $\lambda = JD - 2'458'000$ . In order to isolate these segments, we first identified local minima and maxima along the light curves, while leaving out local minima and maxima caused by a single point, since this would mean fitting a power-law to only two points, which is not statistically significant. We identified a segment as periods of time where a local minima was followed by a local maxima, or viceversa. These isolated segments were then fitted, and from the resultant power-laws, we calculated the doubling/halving times, by first evaluating the power-law on the initial point of each segment, and then solving again for double or half the flux of the same initial point. The fitting to the segments and the doubling/halving times can be found in Figure \ref{fig:flares} and Table \ref{tab:flares}, respectively.

\begin{table*}
\centering
\caption{Doubling/halving times, $t_d$, in the K and I filters. The segments correspond to those in Figure \ref{fig:flares}.  \label{tab:flares}}
\begin{tabular}{lcc}
\hline
Filter & Segment & $t_d$ (days) \\ \hline
K & 1 & 28.1 \\
K & 2 & 6.7 \\
K & 3 & 31.8 \\
K & 4 & 25.6 \\
K & 5 & 34.2 \\
K & 6 & 21.6 \\
I & 1 & 18.2 \\
I & 2 & 12.2 \\
I & 3 & 28.7 \\
I & 4 & 7.7 \\
I & 5 & 35.6 \\
I & 6 & 20.7 \\
I & 7 & 21.5 \\
I & 8 & 20.9 \\
I & 9 & 19.7 \\
I & 10 & 13.6 \\
\hline
\end{tabular}
\end{table*}

\bsp	
\label{lastpage}
\end{document}